\documentclass[12pt,preprint]{aastex}
\usepackage{lineno}
\usepackage{amssymb}
\usepackage{graphicx}
\usepackage{subfigure}
\newcommand{\ww}[1]{{\color{red}[WW: #1]}}
\shorttitle{\ww{The SAGES project -- I. Description and First Data Release}}
\shortauthors{Z. Fan, G. Zhao, W. Wang, et al.}

\begin{document}
\title{The Stellar Abundances and Galactic Evolution Survey (SAGES) -- --
  I. General Description and the First Data Release (DR1)}

\author{Zhou Fan\altaffilmark{1}, Gang Zhao\altaffilmark{1,2},  Wei
  Wang\altaffilmark{1}, Jie Zheng\altaffilmark{1},  Jingkun Zhao\altaffilmark{1}, Chun
  Li\altaffilmark{1}, Yuqin Chen\altaffilmark{1}, Haibo Yuan\altaffilmark{3},  Haining 
  Li\altaffilmark{1}, Kefeng Tan\altaffilmark{1}, Yihan Song\altaffilmark{1}, Fang
  Zuo\altaffilmark{1}, Yang Huang\altaffilmark{1,2}, Ali Luo\altaffilmark{1}, Ali
  Esamdin\altaffilmark{4}, Lu Ma\altaffilmark{4}, Bin 
  Li\altaffilmark{7,8}, Nan Song\altaffilmark{9}, Frank
  Grupp\altaffilmark{10}, Hai-bin Zhao\altaffilmark{7,8},
  Shuhrat.A. Ehgamberdiev\altaffilmark{5,6},
  Otabek. A. Burkhonov\altaffilmark{5},  Guojie
  Feng\altaffilmark{4}, Chunhai Bai\altaffilmark{4}, Xuan
  Zhang\altaffilmark{4},  Hubiao Niu\altaffilmark{4},
  Alisher.S. Khodjaev\altaffilmark{5}, Bakhodir.M. Khafizov\altaffilmark{5},
  Ildar.M. Asfandiyarov\altaffilmark{5},  Asadulla.M. Shaymanov\altaffilmark{5},
  Rivkat.G. Karimov\altaffilmark{5}, Qudratillo. Yuldashev\altaffilmark{5}, 
  Hao Lu\altaffilmark{7}, Getu Zhaori\altaffilmark{7}, Renquan
  Hong\altaffilmark{7}, Longfei Hu\altaffilmark{7}, Yujuan
  Liu\altaffilmark{1}, Zhijian Xu\altaffilmark{7,11}} 

\altaffiltext{1}{CAS Key Laboratory of Optical Astronomy, National
  Astronomical Observatories, Chinese Academy of Sciences, Beijing
  100101, China}

\altaffiltext{2}{School of Astronomy and Space Science, University of
  Chinese Academy of Sciences, Beijing 100049, China}

\altaffiltext{3}{Department of Astronomy, Beijing Normal University,
  Beijing 100875, China} 
  
\altaffiltext{4}{Xinjiang Astronomical Observatory, Urumqi, Xinjiang 830011, China}   

\altaffiltext{5}{Ulugh Beg Astronomical Institute of the Uzbekistan Academy of Sciences, 
  Astronomicheskaya str. 33, Tashkent, Uzbekistan}

\altaffiltext{6}{National University of Uzbekistan, Tashkent, Uzbekistan}

\altaffiltext{7}{Key Laboratory of Planetary Sciences, Purple Mountain
  Observatory, CAS, Nanjing 210023} 

\altaffiltext{8}{School of Astronomy and Space Science, University of
  Science and Technology of China, Hefei, 230026, China} 

\altaffiltext{9}{China Science and Technology Museum, Beijing,
  100101, China}

\altaffiltext{10}{Max Planck Institute for extraterrestrial Physics,
  85748 Garching, Germany} 

\altaffiltext{11}{China Three Gorges University, Yichang 443002, China}

\email{gzhao@bao.ac.cn}

\begin{abstract}
The Stellar Abundances and Galactic Evolution Survey (SAGES) of the
northern sky is a specifically-designed multi-band photometric survey
aiming to provide reliable stellar parameters with accuracy comparable
to those from low-resolution optical spectra. It was carried out with
the 2.3-m Bok telescope of Steward Observatory and three other telescopes.
The observations in the $u_s$ and $v_s$ passband produced 
over 36,092 frames of images in total, covering a sky area of
$\sim9960$ degree$^2$. The median survey completeness of all
observing fields for the two bands are of $u_{\rm s}=20.4$ mag and 
$v_s=20.3$ mag, respectively, while the limiting magnitudes with
signal-to-noise ratio (S/N) of 100 are $u_s\sim17$ mag and $v_s\sim18$
mag, correspondingly. We combined our catalog with the data release 1
(DR1) of the first of Panoramic Survey Telescope And Rapid  
Response System (Pan-STARRS1, PS1) catalog, and obtained a total of
48,553,987 sources which have at least one photometric measurement in each
of the SAGES $u_s$ and $v_s$ and PS1 $grizy$ passbands, which
  is the DR1 of SAGES and it will be released in our paper. We compare our
$gri$ point-source photometry with those of PS1 and found an RMS scatter of
$\sim2$\% in difference of PS1 and SAGES for the same band. We estimated an
internal photometric precision of SAGES to be on the order of
$\sim1$\%. Astrometric precision is better than $0^{\prime 
  \prime}.2$ based on comparison with the DR1 of Gaia mission. In this
paper, we also describe the final end-user database, and provide some
science applications.    
\end{abstract}


\section{Introduction}
\label{intro.sec}

Taking images of the entire or a large portion of digital sky
  surveys, such as the Sloan Digital Sky Survey
\citep[SDSS,][]{york}, which have revolutionized the speed and ease of
making new discoveries. Instrumental limitations imply that surveys 
need to strike a balance between depth and areal coverage. The SDSS,
e.g., imaged 70\% of the Northern hemisphere sky, is one with the
largest sky-coverage surveys. The median $5-\sigma$ depth for SDSS
  photometric observations is $u=22.15$, $g=23.13$, $r=22.70$, 
$i=22.20$, and $z=20.71$, the field of view is 6
deg$^2$ \footnote{https://www.sdss4.org/dr17/imaging/other\_info/.}
while its survey depths, particularly in the $u$-band, are not deep enough,
especially for the purpose of stellar parameters estimation. 

For the photometric survey, the SkyMapper is a southern sky
survey project led by Australian National University \citep{kell}. Its
scientific targets include objects in 
the solar system, the formation history of young stars in the solar
neighborhood, the distribution of dark matter halo in the Milky Way,
atmospheric parameters of $\sim$100 million stars, extremely metal
poor stars, photometry redshift calibration of galaxies, and
high-redshift quasars. The survey was performed using the Siding
Spring Observatory's 1.35-m telescope. The telescope has a large field
of view of $2.3\times2.3$ square degrees and can perform observations in six
bands ($u$, $v$, $g$, $r$, $i$, $z$), with a limiting magnitude of 
$u_{\rm SM}\sim16.8$ mag, $v_{\rm SM}\sim17.0$ mag (10$\sigma$). 
For the spectroscopic survey, the Large Sky Area Multi-Object
Fibre Spectroscopic Telescope   
(LAMOST) Data Release 7 (DR7, http://dr7.lamost.org/) has already released 
more than 14 million stellar spectra, including 10.4 million (73.3 \%) low
resolution and 3.8 million medium resolution (the rest 26.7\%), which
provides the largest database of stellar spectra at present. In
addition, DR7 also provides a large stellar parameter catalog in
the world (6.91 million).  
The Gaia space mission \citep{2016A&A...595A...1G} mainly propose to
provide high-precision astrometry of point sources; furthermore, their
BP/RP spectra and RVS spectra have been used to estimate stellar
parameters of 470 million and 6 million stars respectively as well
{\bf \citep{rb22}}. The Euclid Wide
Survey \citep[EWS,][]{ECoI} also has the wide area imaging and
slitless spectroscopy instruments in the optical and NIR band, which
could be used for the study of the Milky Way.

The Str$\rm\ddot{o}$mgren-Crawford (hereafter SC) system is a 6-color
medium and narrowband photometric system. It was first introduced 
by \cite{str56} containing the four intermediate bands $uvby$ and
was then supplemented with the two H$\beta$ filters
($\beta=\beta_w-\beta_ n$) by
\cite{cra58}. The SC system includes four color indices,
namely $b-y$, $m1$, $c1$ and $\beta$, which are widely used for the
effective and reasonably accurate determination of stellar atmospheric
parameters including $T_{\rm eff}$, log~$g$ and [Fe/H], etc. 
The SC system is originally designed for the study of A2-G2 type stars
\citep{str63a}, later it was found to be able to provide
useful constraints on other various types of targets, including giant
stars \citep{anth}, red giant branch stars \citep{gust}, yellow
supergiants \citep{are}, G and K dwarf stars \citep{twa}, metal-poor
stars \citep{shu}, metal-poor giants\citep{cham}. Other 
advantages of the SC system include determining the stage of stellar
evolution \citep{str63b} and estimating the interstellar reddening and
extinction \citep{pau15}.  When extinction is known, the stellar
distance can be estimated once the evolutionary stage is
determined. The SC photometric system needs relatively long
integration exposure and thus more consumption in time and manpower,
due to narrower bandwidths as compared to broadband photometric surveys. 

The two most widely-used data of the SC system sky survey so far are
the Geneva-Copenhagen survey of the Solar neighborhood (GCS) \citep{nord} 
and HM sky survey \citep{hm}. In fact, they are not “real” sky
surveys, but two catalogs based on collecting and collating the
literature data obtained in the SC system. For example, the GCS Sky Survey 
selected 30,465 stars whose apparent magnitude $M_V$ brighter than
8.5 mag from the Henry Draper (HD) catalog \citep{cp18a, cp18b} and
complete to a distance of $\sim$40 pc, mostly in 
the solar neighborhood. Due to some color limitations, they finally
selected an unbiased dynamic sample of 16,682 stars with relatively
uniform spatial distribution and complete volume. Many studies of the
solar neighborhood and the Galaxy are based on this sample. The HM
survey simply collects 63,313 stars whose $u$, $v$, $b$,
$y$, $\beta$ photometric data have been obtained. The magnitude distribution
of the $y$-band (the central wavelength is close to the $V$-band) is 
$\sim5-15$ mag, approximately a Gaussian distribution. The mean value
and standard deviation are 8.41 mag and 1.81 mag, respectively. 
The sample is complete down to $\sim8-9$ mag in these bands, 
 according to the distribution of the magnitude.
 In terms of sample size, survey depth and sky coverage, the
current available intermediate to narrow-band photometric data is far
from enough for the study of our Galaxy (due to the disadvantage of
longer exposure times /or less depth), thus calling for a much 
deeper and wider-area sky survey. 

The recent Stellar Abundances and Galactic Evolution Survey (SAGES)
\citep[PI: Gang Zhao,][]{fan18,zheng18,zheng19} is the one
for such a purpose. The survey aims to derive stellar atmospheric
parameters for a few hundred million stars in the
$u_s$/$v_s$/g/r/i/$\alpha_n$/$\alpha_w$/DDO51 bands. It plans to  
cover the northern sky region of Dec $\delta>-5$ deg 
and avoid the Galactic disk region of $-10$ deg $< b < +10$ deg
  while avoiding
saturation and image contamination that may result from excessive
bright stars. In addition, we selected the region of $\alpha>$ 18 h
and $\alpha<$ 12 h for RA to be the first-prioritized survey area,
which can be accessed in autumn and winter, the best observing seasons
at the Kitt Peak National Observatory (KPNO).  The final survey area
was larger than 12,000 square degrees (see
Section~\ref{obs.sec} for details). In order to facilitate flux calibration 
between images, 20\% overlap is reserved for each sky field and all
adjacent fields. The SAGES has limiting magnitude (S/N$\sim 100$)
of $u_s \approx 17.5$ mag, $v_s \approx 16.5$ mag, about $\sim 8-9$
mag deeper than the GCS and HM surveys. This corresponds a
completeness distance of $\sim1$ kpc for a solar-like star
\citep{fan18} and $\sim25$ times as large as that of GCS.
Currently, SAGES has already covered $\sim9960$ deg$^2$ of
the northern sky in the $u_s$ and $v_s$ bands, almost
completing the original survey plan. Its $v_s$-band filter is designed by
the SAGES team covering the Ca II H\&K lines, aiming to provide
reliable stellar metallicity, while $u_s$ (similar to the $u$ band
of Stromgren-Crawford system, covering the Balmer jump) could provide
constraints on surface gravity for at least early type stars. The data
can be used to derive the age and metallicity of stellar populations in the Milky
Way, and nearby galaxies including M31/M33, as well as precise
interstellar extinctions of individual stars. 

In this paper, we introduced the SAGES project and present the 
first data release (DR1). This is organized as follows. 
Section~\ref{sur.sec} describes the survey details, including the
design of the SAGES photometric system, telescopes and
instruments, the observing strategy and sky coverage. In Section~\ref{pip.sec}
we show the observations of the SAGES; 
Section~\ref{dat.sec} describes the data product and the release;
Section~\ref{sci.sec} presents a few potential science cases that may
be conducted using the SAGES data, followed by a future
prospect in Section~\ref{fut.sec}.

\section{The Description of SAGES}
\label{sur.sec}

\subsection{The Survey and Operations}

The SAGES project is an international cooperative survey project which
utilized four survey telescope facilities worldwide. We aim to observe 
a large part of the northern sky (except the sky area of Galactic plane, $|b|
>10$ degree, $\rm >12000~degree^2$, 58.2\% of the northern sky) and 
obtained a large intermediate-band or narrow-band photometric catalog of stars 
which are much deeper ($\sim8-9 $ mag) than that of GCS and HM
  of the limiting magnitude for constraining the stellar
  parameters. Thus we can  
derive the stellar parameters from our SAGES catalog and the magnitude 
coverage can be well combined for the SAGES faint end and GCS bright end.
Nightly survey operations are planned by autonomous
scheduler software that can execute the entire survey without 
human intervention.  

\subsection{The Filter Design and SAGES Photometric System}
\label{fil.sec}

 In our survey, we will carry out the observations in
  $u_s$/$v_s$/g/r/i/$\alpha_n$/$\alpha_w$/DDO51 bands.
The filters are chosen for the following reasons. We aim to
derive the stellar parameters for a large sample of northern sky
with photometry, which is similar to the SC  
system. As mentioned in Section~\ref{intro.sec}, the SAGES $u_s$ is the 
similar $u$ passband of Stromgren-Crawford (SC) system, which covers the 
Balmer jump, and it is sensitive to the stellar photospheric gravity; the
$v_{\rm s}$-band filter is designed by ourselves and covers the Ca II H\&K
absorption lines which are very sensitive to stellar metallicity. 
The $gri$ bands are the same as the SDSS passbands, which are
used to estimate the effective temperature $T_{\rm eff}$. 
The intermediate band DDO51 measures the MgH feature in KM dwarfs
  \citep{bess05}, which is sensitive to gravity of late type stars. 
The other two bands, $\alpha_w$ and $\alpha_n$, are used
to estimate the interstellar extinctions, as the value of
$\alpha=\alpha_w-\alpha_n$, which is designed by ourselves and
  similar to the 
$\beta=\beta_w-\beta_ n$  in the SC system
\citep{cra58}. It is only sensitive to the effective 
temperature $T_{\rm eff}$, and it is independent of interstellar
extinction. Thus the photometry of the two passbands can be used to
estimate the interstellar extinction. 
However, for the CCD photometry, the quantum efficiency (QE) is
much higher in the wavelength  
of the $\alpha_w$ and the $\alpha_w$ absorption line is stronger for
the FGK stars. We will describe the advantages and sensitivity
tests in the following.

Table~\ref{t1.tab} shows the central wavelength and bandwidths of the
filters of the SAGES photometric system. We used the prime focus system of the
2.3-m (90-inch) Bok telescope for observations in the $u_s$ and $v_s$ passbands.
The Bok telescope belongs to the Steward Observatory, University of Arizona 
which is located at KPNO. 

We ordered the SAGES filters from different manufactures:
for the $u_s$ band filter on the 90prime telescope, it was made in the Omega
Optical Inc, USA; for the $v_s$ filter on the 90prime telescope, it was made
in Asahi Spectra Co., Ltd, Japan, which has very high efficiency; for the $\alpha_w$ and
DDO51 on the Xuyi 1-m telescope, they were made in Beijing Bodian Optical Technology
Co. Ltd;  for the $\alpha_w$ and $\alpha_n$ and DDO51 filters of MAO 1-m
telescope, they were made in Asahi Spectra Co., Ltd, Japan, which also
have very high efficiency; for the $gri$ band filters on 1-m telescope of
Nanshan station of XAO, which were made in Custom Scientific, Inc,
USA, which are the standard SDSS system.

\pagestyle{empty}
\begin{deluxetable}{lcc}
  \tablecolumns{3} \tablewidth{0pc} \tablecaption{The $u_s$ and $v_s$ passband 
  filters of SAGES: properties.  
    \label{t1.tab}}
  \tablehead{
    \colhead{Bandpass} & \colhead{$u_s$} &  \colhead{$v_s$} \\
    }
  \startdata
Central Wavelength (\AA) & 3425 & 3950 \\ 
Bandwidth (\AA) & 314 & 290 \\ 
  \enddata
\end{deluxetable}

\begin{figure}
  \centering
  \includegraphics[angle=0,scale=0.4]{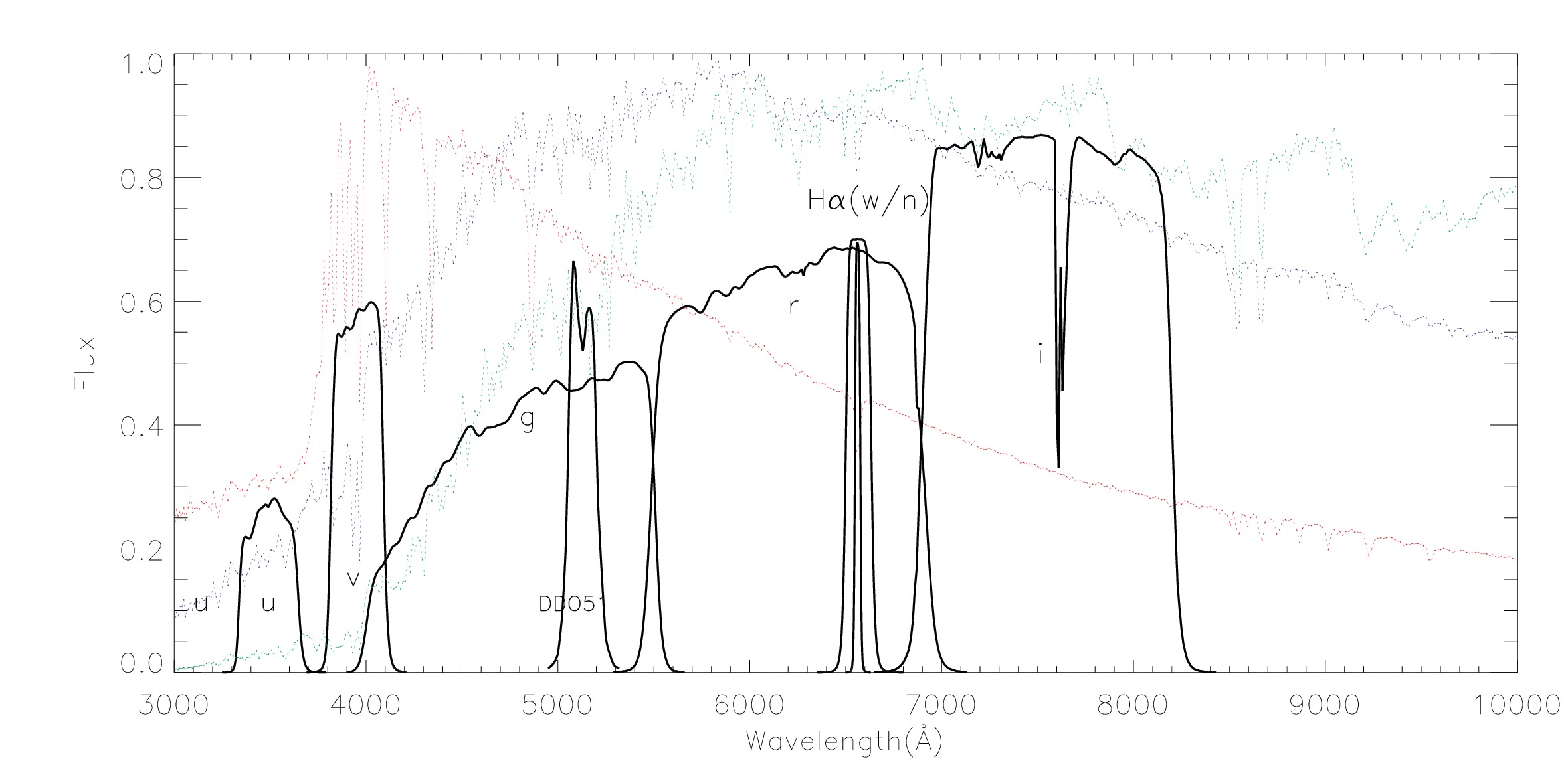}
  \caption{Spectra of typical F- (red) G- (blue) K- (green) type
    stars from MILES overplotted with the SAGES filter passbands.} 
  \label{fig1}
\end{figure}

Since the FGK- type of stars are good tracers to study the 
nature of Milky Way, we focus on the stellar parameters of the FGK-type of stars.
Figure~\ref{fig1} shows the spectra of the MILES\footnote{
https://lco.global/$\sim$apickles/INGS/ingsColors.supdat} of the FGK
stars with the filter transmission of the SAGES 
passbands for the FGK- type stars. We see that it is easy to distinguish 
the different types of stars with the color of the SAGES photometry.

\begin{figure}
  \centering
  \includegraphics[angle=0,scale=0.8]{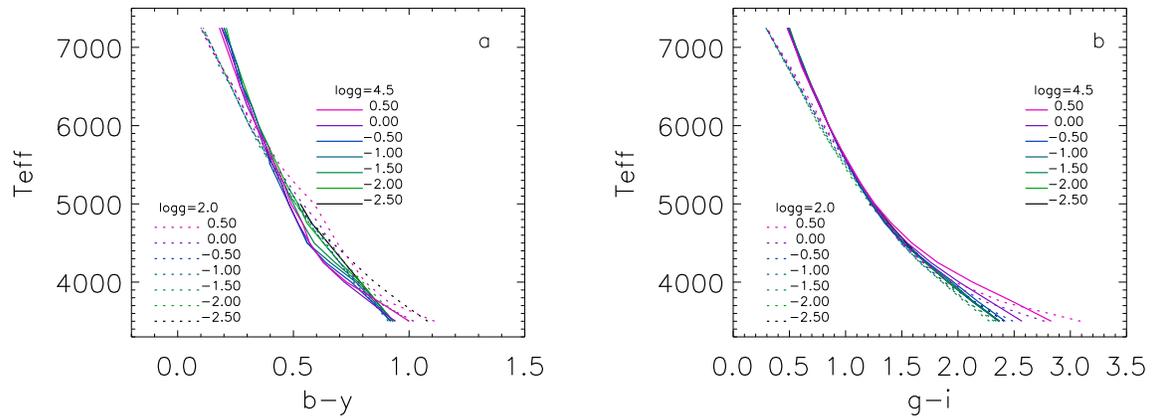}
  \caption{The effective temperature versus colors for
    different metallicities with two given gravities of log~$g=2.0$ (a) and 4.5 (b).}
  \label{fig2}
\end{figure}

The advantage of the SAGES filter system: which are more
sensitive to stellar parameters than that of the traditional SC filter
system. We utilized the Kurucz model \citep{mun05} to analyze the dependence of the
the colors of SAGES and the stellar parameters. We focus on the analysis
of the gravity values log~$g=2$ and 4.5, and the metallicity range of
$\rm [Fe/H]=-2.5$ to 0.5.  Figure~\ref{fig2} shows the effective
temperature versus colors for different metallicities with two given
gravity log~$g=2.0$ (a) and 4.5 (b). It shows that the SAGES system is clearly
much more sensitive to the effective temperature than that of the SC system.
We can see that for the same effective temperature range, no matter
which type of stars (FGK), the color range of SAGES is $\sim2-3$ times 
of that in the SC system. In this case, the uncertainty of the effective
temperature $T_{\rm eff}$ is 1/3 to 1/2 times of that in the SC system,
improving the accuracy by $\sim2-3$ times, i.e., the
uncertainty of the effective temperature $T_{\rm eff}$ of a
star of $V=15$ mag in the SC system is comparable to uncertainty 
$V\sim16-16.5$ mag for the same star in the SAGES system \citep{fan18}.

\begin{figure}
  \centering
  \includegraphics[angle=0,scale=0.8]{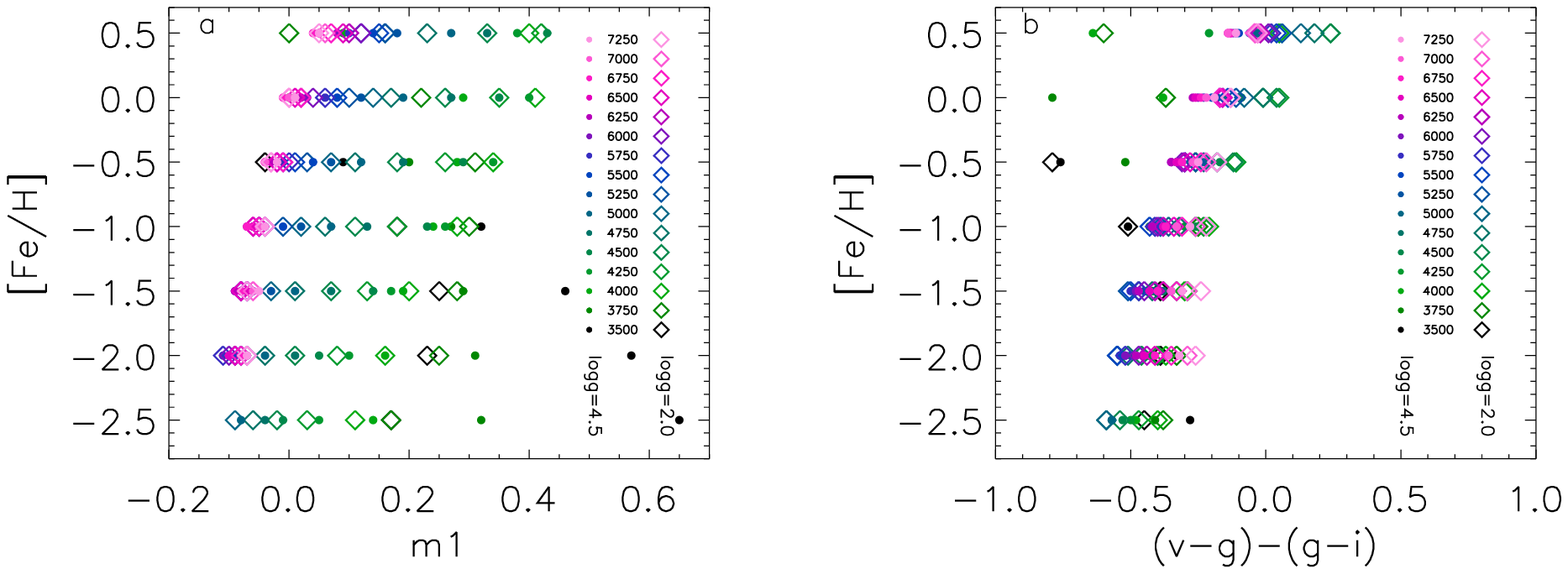}
  \caption{Metallicity versus colors of the SC system (left panel)
    and SAGES (right panel) for different
    effective temperatures with a given gravity log~$g=2.0$ and 4.5.} 
  \label{fig3}
\end{figure}

Further, we have also investigated the relationship between the metallicity and
colors of SC system (left panel) and SAGES (right
panel). Figure~\ref{fig3} shows the relations between 
metallicity and colors of the SAGES for different effective
temperatures with a given gravity of log~$g=2.0$ and 4.5. Clearly, the SAGES
system has higher sensitivity of the metallicity than that of the SC system.
We can see that for the same metallicity range, no matter
which type of stars (FGK), the color range of SAGES is $\sim2-4$ times
that of the SC system. In this case, the uncertainty of the metallicity
$\rm [Fe/H]$ is 1/4 to 1/2 times of that in the SC system, improving the
accuracy by $\sim2-4$ times, i.e., the uncertainty of the metallicity
$\rm [Fe/H]$  for a star of $V=15$ mag in the SC system is comparable to
that of $v_s\sim16-17$ mag for the same star in the SAGES system, which can be
seen from the spectra of FGK stars shown in Figure~\ref{fig1}.

\begin{figure}
  \centering
  \includegraphics[angle=0,scale=0.8]{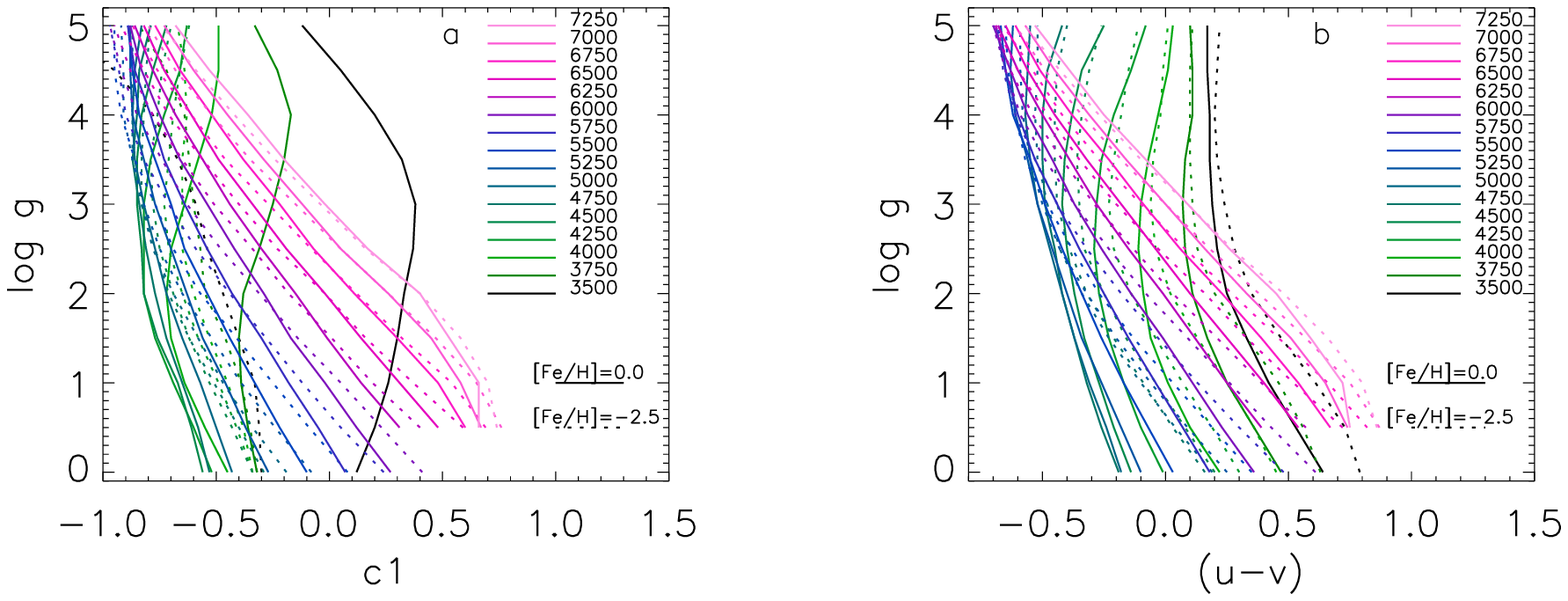}
  \caption{Gravity versus colors of the SC system (left panel) and
    SAGES (right panel) for different effective temperatures with solar metallicity.} 
  \label{fig4}
\end{figure}

Figure~\ref{fig4} presents the relations between the gravity and colors of the
SC system and SAGES for different effective temperatures. We consider two  
cases where the metallicity varies widely, e.g., for the metallicity $\rm
[Fe/H]=-2.5$ and 0. For FG-type stars, the SAGE system is slightly
more sensitive than the SC system (as shown in Figure~\ref{fig4}). For
K-type stars, the SC system has a serious "S - shape" i.e., the
non-monotonic relation. Although the SAGES system also shows a
non-monotonic condition, the relationship between gravity and color is
a monotonic function in the interval between log~$g \gtrsim 2$ and log~$g 
\lesssim 2$ if the distinction is made at log~$g\approx2$. We use the DDO51 filter, 
which can effectively distinguish dwarf stars and giant stars, to provide a
feasible solution of gravity log~$g$ for K-type stars, which is
clearly the advantage of the SAGES photometric system \citep{fan18}.

\begin{figure}
  \centering
 \includegraphics[angle=0,width=0.9\textwidth]{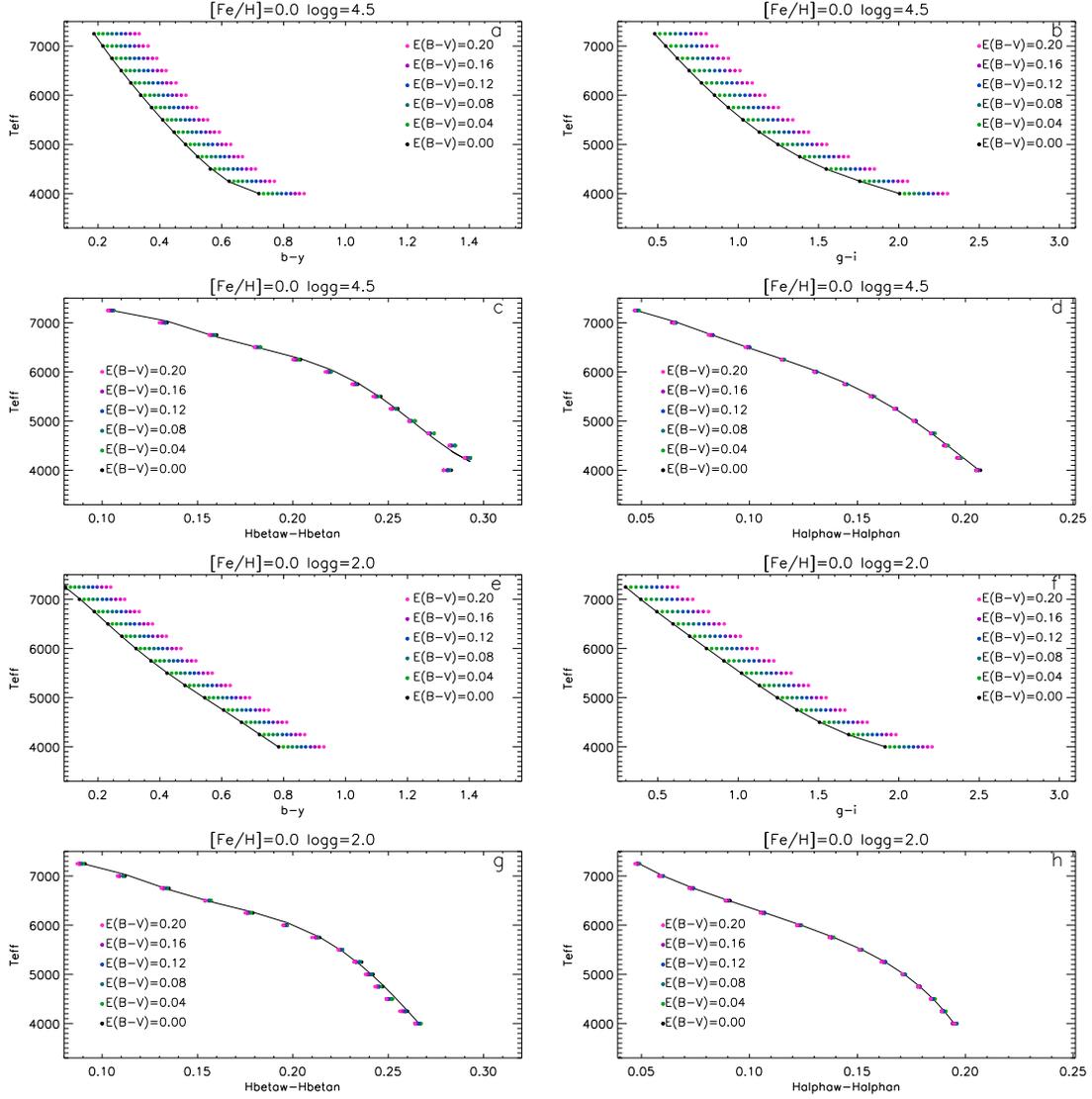}
  \caption{The extinction variations in the effective temperature
    $T_{\rm eff}$ versus colors of the SAGES for different log~$g$ with solar
    metallicity and comparisons of the $\beta_n-\beta_w$ and
    $\alpha_n-\alpha_w$ for the extinction variations. The gravity
    log~$g=4.5$ for the panels of a, b, c and d, while log~$g= 2.0$
    for the panels of e, f, g and h; a, c, e and g panels are
    for the SC system and b, d, f and e panels are for the SAGES.} 
  \label{fig5}
\end{figure}

Figure~\ref{fig5} shows the relation between the effective temperature $T_{\rm
  eff}$ and colors of SAGES for different gravity log~$g$ with 
solar metallicity, where log~$g=4.5$ for the panels of a, b, c and d; log~$g=2.0$ for 
for the panels of e, f, g and h. For comparison, a, c, e and g 
panels are for the SC system and b, d, f and e panels are for SAGES.
For G- and K-type stars, the absorption line strength of H$\beta$ is weaker
than that of H$\alpha$, proving the intensity measurement accuracy of
H$\alpha$ is higher. The relationship between the color index and different
extinctions can be obtained by calculation. We take log~$g=2$ and 4.5
and $\rm [Fe/H]= 0$. Figure~\ref{fig5} shows the 
extinction variations in the relations between effective 
temperature and color: $g-I$ is more sensitive to effective
temperature than $b-y$; For different extinction,
$\beta_n-\beta_w$ has a certain color change ($\sim0.01$ mag),
while for $\alpha_n-\alpha_w$, the change with extinction is
almost invisible (E(B$-$V) ranges from 0 to 0.2), suggesting that the
latter is more independent of interstellar extinction. Thus the solution
will be more accurate \citep{fan18} for $\alpha_n-\alpha_w$
than that for $\beta_n-\beta_w$.  

\subsection{Telescopes and Detectors of SAGES}

\begin{figure}
  \centering
 \includegraphics[angle=0,scale=0.3]{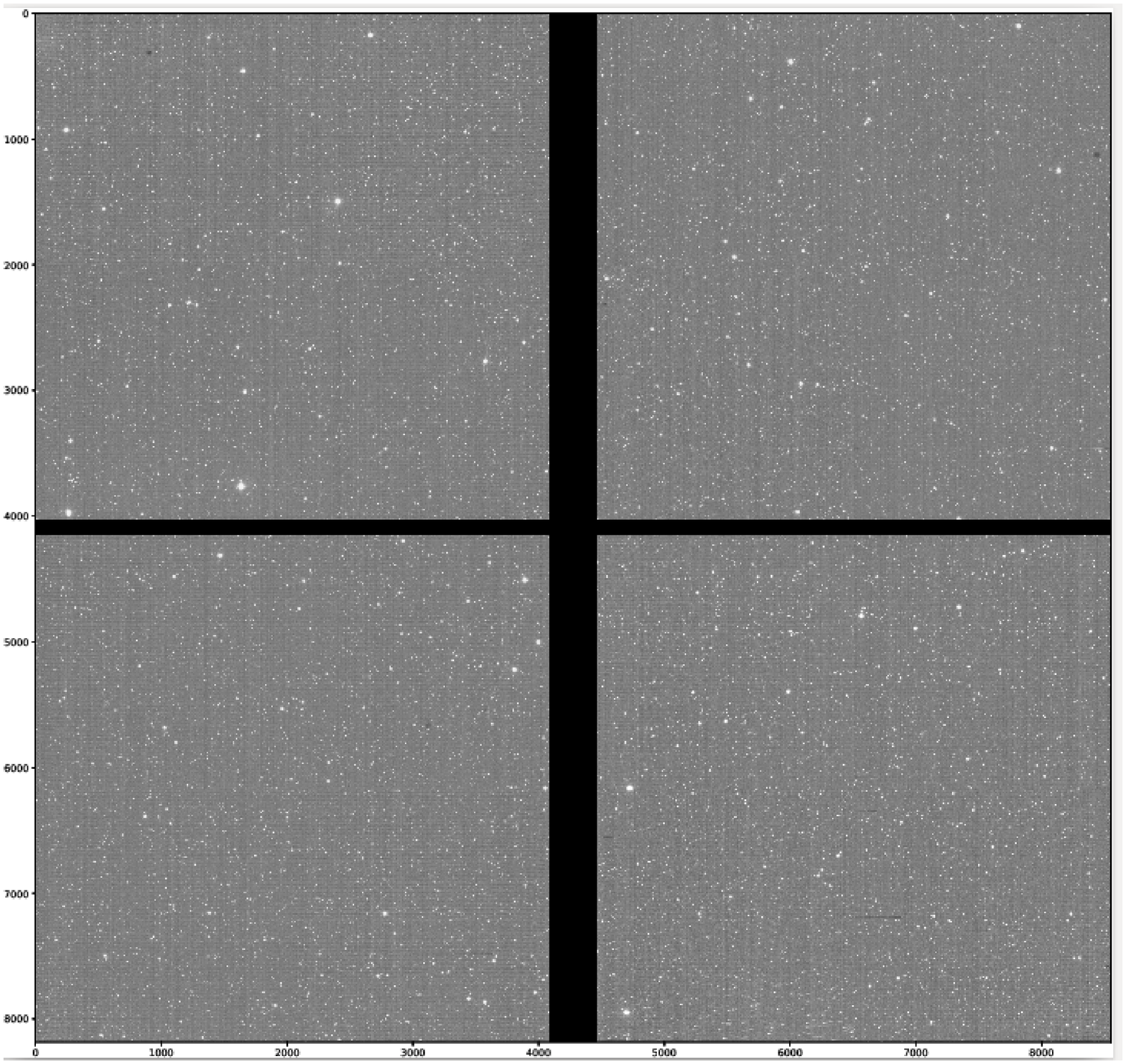}
  \caption{The image of the Bok telesocpe in the SAGES $v_s$ band. The CCD
    mosaic is composed of four CCDs and for each one there are four amplifiers.}
  \label{fig6}
\end{figure}

For the $u_s$- and $v_s$-passband observations were carried out with
the 90-inch (2.3-m) Bok telescope in Kitt Peak of Arizona,  USA. The
prime focus is adopted for the survey, with 
corrected focal ratio of f/2.98 and corrected focal length of
6829.2 mm. The altitude of Kitt Peak National Observatory  
(KPNO, 111$^{\circ}$36'01''.6W, +30$^{\circ}$57'46''.5N) is 2071
meter. The location offers very stable seeing conditions and a fairly  
low horizon in all directions save for the northeast. 
The sky brightness measurements of Kitt Peak in 2009 are 22.79
  mag arcsec$^{-2}$ in B and between 21.95 mag arcsec$^{-2}$ in V band
  \citep{nm10}.  For the location
of the Bok telescope, the typical seeing was $1''.5$ during the
observations. For the Bok 
telescope, an 8K $\times$ 8K CCD mosaic is composed of four 4K
$\times$ 4K blue optimized sensitive back luminous CCDs, with gaps
along both the RA (166$^{\prime\prime}$) and Dec directions
(54$^{\prime\prime}$), by the University of 
Arizona Imaging Technology Laboratory (ITL). In figure~\ref{fig6}, we
can see the distribution of the four CCD arrays at the prime focus of
the Bok telescope. The QE is $\sim$80 \%  
for the u-band (central wavelength of $\sim3538$ {\AA} with FWHM of
520 \AA) and the edge-to-edge field of view (FOV) is about
$1^{\circ}. 08\times1^{\circ}. 03$ \citep{zou15, zou16, zhou16}. In
this paper, only the observing data of 2.3-m Bok telescope are released.

The observing data of the following three telescopes are not
included in this paper and will be released in following works:

For the $gri$ passband, NOWT of Xinjiang Observatory, CAS;  A one-meter 
wide field astronomical telescope with Alt-Az mount, operating at prime 
focus with a field corrector. The NOWT provides excellent optical quality, 
pointing accuracy and tracking accuracy. It is located at Nanshan
(87$^\circ$10.67$^{\prime}$E, 43$^\circ$28.27$^{\prime}$ N) with an
altitude of 2080~m, which is $\sim$75~km away from Urumqi city. A 4K
$\times$ 4K CCD was  
mounted on the prime focus of the telescope. The FOV at prime focus is 
$1.5^\circ \times 1.5^\circ$, and pointing accuracy is better than 
$5^{\prime\prime}$ RMS for each axis after pointing model correction \citep{liu13}. 

For the $\alpha_n$ passband and part of $\alpha_w$ passband,
1-m Zeiss telescope of Maidanak astronomical observatory (MAO,
66$^\circ$53'47''E, 38$^\circ$40' 22''N), which 
belongs to Ulugh Beg Astronomical Institute (UBAI), Uzbekistan. UBAI
is an observational facility of the Uzbekistan Academy of
Sciences (UAS). The total amount of clear night time is 2000 hours with
median seeing of 0$^{\prime\prime}$.70. The altitude is 2593~meter.  
Based on two month observation performed in 1976,
it shows that the sky background of Mount Maidanak varies between $22.3
- 22.9$ mag arcsec$^{-2}$  in B and between $21.4 - 22.0$ mag arcsec$^{-2}$
in V band \citep{kf76}. Its current field
of view is $32.9' \times 32.9'$. For the narrow-band photometry, if the
focal ratio is too fast, the central wavelength and the bandwidth will
shift to some extent. For its Cassegrain focus, the focal ratio is f/13. However,
in order to enlarge the FOV, we installed a focal reducer to make the
focal ratio f/6.5, which is still suitable for the narrow-band survey. 
A 2K $\times$ 2K Andor DZ936 CCD was mounted on the f/6.5 Cassegrain focus 
of the telescope, to ensure that there is no wavelength shift or bandwidth 
changes for the narrow-band filters \citep{ehga00,ehga18}.

For the $\alpha_w$ and DDO51 band, Xuyi 1-m Schmidt 
telescope of Purple Mountain Observatory (PMO) of CAS: The Xuyi Schmidt
Telescope is a traditional ground-based refractive-reflective
telescope with a diameter of 1.04/1.20 meter. It is equipped with a
10K $\times$ 10K thinned CCD camera, yielding a $3.02^\circ \times 3.02^\circ$ 
effective FOV at a sampling of $1^{\prime\prime}.03$ per pixel projected on
the sky. The QE of the CCD, at the cooled working temperature
of $-103.45^\circ$ C, has a peak value of 90\% in the blue and remains
above 70\% even to wavelengths as long as 8000 {\rm \AA} . The
XSTPS-GAC was carried out with the SDSS $g$, $r$ and $i$ filters. The
current work presents measurements of the Xuyi atmospheric
extinction coefficients and the night sky brightness in the three SDSS
filters based on the images collected by the XSTPS-GAC. The night sky
brightness determined from images with good quality has median values
of 21.7, 20.8 and 20.0 mag arcsec$^{-2}$ and reaches 22.1, 21.2 and 20.4
mag arcsec$^{-2}$ under the best observing conditions for the $g$, $r$ and $i$
bands, respectively \citep{zhang13}. The typical limiting magnitude
is of $m(r) \sim20.0$ mag. 

\subsection{Observing Strategy and Coverage}
\label{obs.sec}

\begin{figure}
  \centering
  \includegraphics[angle=0,scale=0.25]{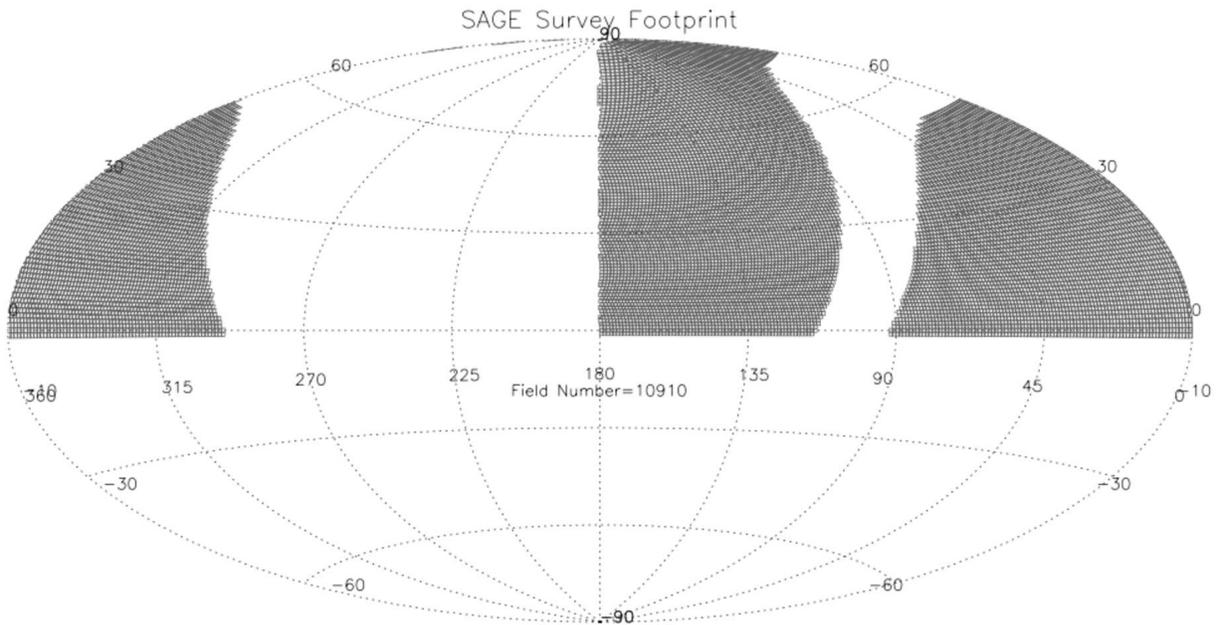}
  \caption{The sky coverage of the SAGES project covers 
most of the northern sky area except the Galactic plane, for which the
Galactic latitude $|b| >10$ degree and the declination is from $-5^{\circ}$ to
$+90^{\circ}$. The total survey area is $\sim$12,000 degree$^2$. }  
\label{fig7}
\end{figure}

Figure~\ref{fig7} shows the survey area (in grey) in equatorial
coordinates. The SAGES covers most of the northern sky
area except the Galactic plane, for which the Galactic latitude $|b|
>10$ degree and the declination is from $-5^{\circ}$ to $+90^{\circ}$. The total
planned survey area $>12,000$ degree$^2$,  which is $\sim60\%$ of the
northern hemisphere.   

\begin{figure}
  \centering
  \includegraphics[angle=0,scale=0.52]{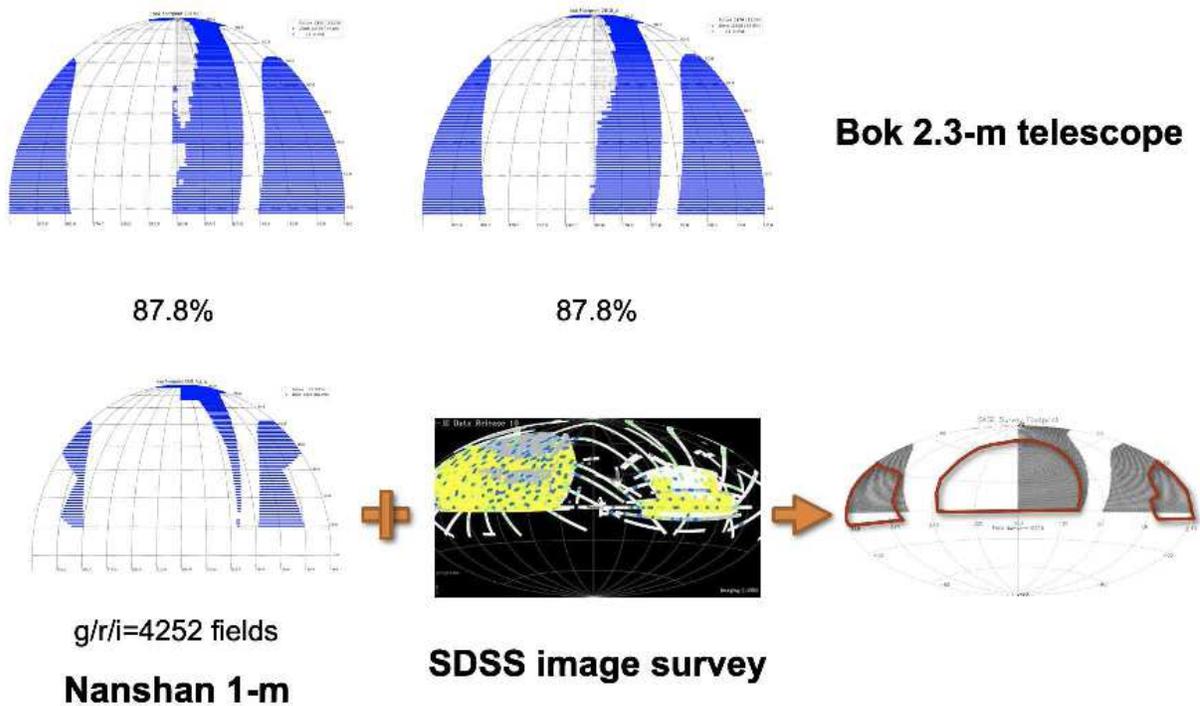}
  \caption{The sky coverage for the SAGES in $u_s$/$v_s$
    passband is shown on the top panel. About
    $87.8\%$ of the fields in total have been observed for 
    SAGES in both passbands, which uses the Bok 2.3m telescope.  
    For the $gri$ passbands (which are shown on the bottom
      panel), we have finished the observations for 
    the designed area with the NOWT, which can be combined with the
    SDSS data for the brighter part of the catalog. The $gri$-band data of NOWT
    will be released in following papers.} 
  \label{fig8}
\end{figure}

{\bf On the top panel}, Figure~\ref{fig8} presents the sky coverage
for SAGES in $u_s$/$v_s$  
observations carried out with the Bok 2.3-m telescope. It covers the
sky area of $\sim9960$ degree$^2$, which is actually $\sim87.8\%$ of
the fields in total observed for SAGES in both passbands. 
Finally we obtained 23,980 frames for $u_s$-band and 17,565 frames for
$v_s$ band. Among these images, 36,092 frames have better quality
while the rest frames, which have been excluded, are defined in the following:
 
1. The frames with failed astrometry in data processing have been 
removed; 

2. For the early testing frames of exposure very short, like 3 seconds
6 seconds, were removed;  

3. No LAMOST standard stars and the number of common sources
with the surrounding celestial region is very small, which also have
been removed.

These "better quality" frames are involved the flux calibration, which
have been released in this paper.

The bottom panel shows that the the $gri$ passbands of SAGES. So
  far we have finished the observations for the designed area with the
  NOWT, which can be combined with the SDSS data for the brighter part
  of the catalog. The $gri$-band data of NOWT will be released in
  following papers.

\begin{figure}
  \centering
  \includegraphics[angle=0,scale=0.45]{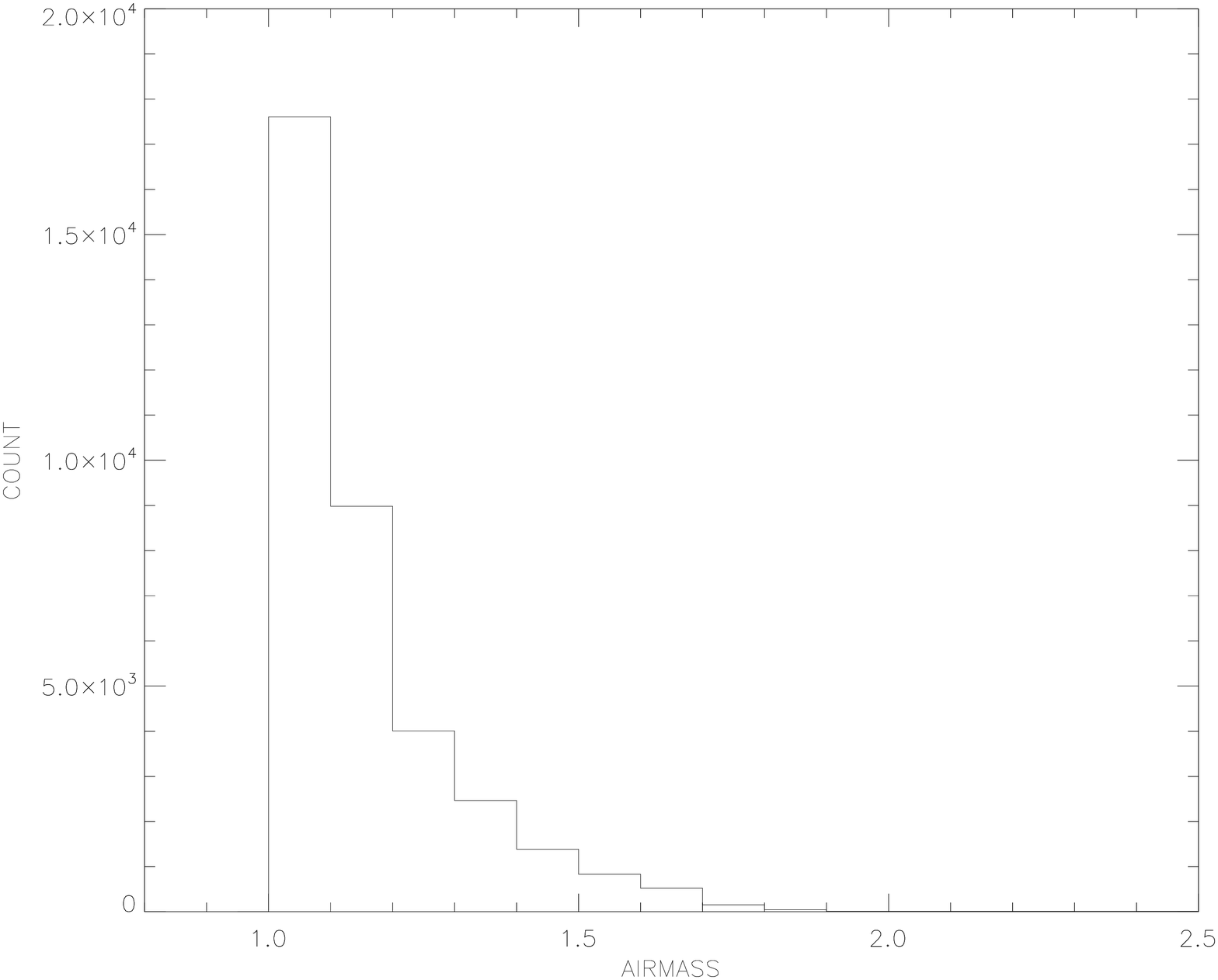}
  \caption{The airmass distribution for SAGES observations, with the
    median value of 1.10. Clearly for most of the fields, the airmass is 
    less than 1.2, which can ensure the quality of our images. This
    aspect has been incorporated in our survey strategy.} 
  \label{fig9}
\end{figure}

Figure~\ref{fig9} shows the airmass distribution for SAGES
observations, with the median value of 1.11, and extending to airmass
of 1.5 as the airmass limit of our observations was set to 1.5 in the
observing strategy. Clearly 
for most of the fields, the airmass is less than 1.2, which can ensure
that our images maintain good quality. This aspect has been
incorporated in our survey strategy. 
    
For the $gri$ passbands, we have completed the observations 
for the designed area with the NOWT of XAO, which can
be combined with the SDSS data. NOWT $gri$+SDSS survey imaging
coverage is the total area for the initial plan of SAGES. The $gri$
band data of NOWT are not included in this paper and will be released
in following papers. However, after the Pan-STARRS DR1 (PS1) data
were released, since the sky coverage is large and the observing is
homogeneous, it is a better choice for us to use the PS1 data of gri
band instead of the NOWT $gri$ +SDSS survey data for the current work.
Another reason is the PS1 can match our SAGES better in the magnitude
range. However, in the future work, the NOWT $gri$ data also can be
used for the stellar parameter estimates.

\section{Data Pipeline}
\label{pip.sec}

\subsection{Image Processing}
\label{ima.sec}

The raw data of SAGES are processed with the pipeline developed by
our data reduction team. The basic corrections of overscan, bias,
flat, and crosstalk effect are the same as the data reduction
pipelines for most similar surveys. 

For the bias correction, we took a set of 10 bias frames before and
after the sky survey observations respectively every observing night.  We 
constructed the combined master bias frame every night by using the
median value of the 20 exposures at the same pixel. Thus both the
  science images and the flat-field images will be corrected for bias
  by applying this master bias image, which are not only corrected for the mean
  value of bias but also for the structure of bias \citep{zheng18,zheng19}.

For the flat-fielding, we took the dome flats with a screen
and a UV lamp to correct the pixel-to-pixel variations. Also we
took the twilight flats to correct large-scale illumination trend if
possible. Otherwise,  a super night-sky flat will be used instead,
which is the combination of all the science images taken over
the night. For most nights, the uncertainty of the master flat images was 
less than 1\%. 

For the highly efficient readout mode can significantly reduce
  the time spent on reading the detector. However, it induces amplifier cross-talk, 
which may cause contamination across the output amplifiers at the
symmetric place, typically at the level of 1:10,000 of the flux. This
effect usually is quite significant for bright saturated stars
on the CCD chips. A large number of images have been used to estimate
the crosstalk coefficients between amplifiers. The overall ratio of
crosstalk is at the level of 5:10,000 for 90Prime, while the inter-CCD 
crosstalk ratios are greater and intra-CCD ratios are lower. 

Figure~\ref{fig6} is one image in the SAGES $v_s$-band of the survey. The
CCD mosaic is composed of four $\rm 4K\times4K$ blue sensitive
back-luminated CCDs. For each CCD, the four-amplifier readout mode is
applied, which has a faster readout speed.

\subsection{Photometry}
\label{phot.sec}

As mentioned above in the Section~\ref{obs.sec}, we obtained 23,980 frames for 
the $u_s$-band and 17,565 frames for the $v_s$ band. In total we have 41,545 
frames for $u_s$ and $v_s$ bands. Among these images, we have 36,092 frames 
with better quality and which involve flux calibration.

In our SAGES pipeline for photometry, we applied the Source Extractor
\citep[SE,][]{ber96} for detecting sources and photometry. The detection
threshold is set to 4 (sigma above the background rms), which
can make sure that most sources could be 
detected and measured precisely. SE could provide the following
measurements and errors: the central positions of each source in both CCD
physical coordinates and celestial coordinates, the roundness and
sharpness, and instrumental magnitudes are included in a series of given 
apertures. 

As we know, for the different photometric methods, the routine will
produce different results. We use the software SE MAG\_AUTO as our
primary output parameters, as this output is in general 
reasonable for both point sources and extended sources. In our
photometric pipeline, the aperture correction needs to be applied to
aperture photometry, which is using the aperture growth-curve method. 

In the SAGES photometric calibration, we adopted the ``AB system''
\citep{og} which is more commonly used for the photometric system, as
it is well-known for the Sloan Digital Sky Survey (SDSS) \citep{york}
by \cite{fuk}.  For our calibration,  the 
comparisons show that the SAGES implementation of the AB system has
an accuracy of $\sim$0.02 mag (90\% confidence). The dominant
contribution is the uncertainty in how well spectrophotometry matches the
AB system. 

\begin{figure}
  \centering
  \includegraphics[angle=0,scale=0.75]{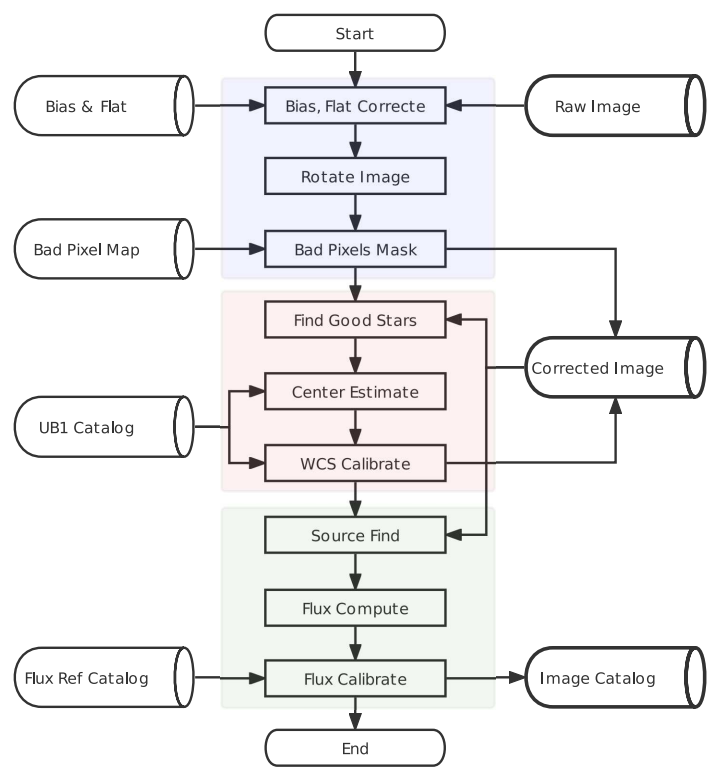}
  \caption{The flow chart of photometric pipeline for SAGES data reduction.}
  \label{fig12}
\end{figure}

Figure~\ref{fig12} shows the flow chart of the photometry pipeline for SAGES, which
uses SE as the main routine for photometry. The
bias and overscan are combined for the correction. We use the
super-flats as final flat for the correction, which is more
reasonable for our data reduction.

\begin{figure}
  \centering
  \includegraphics[angle=0,scale=0.3]{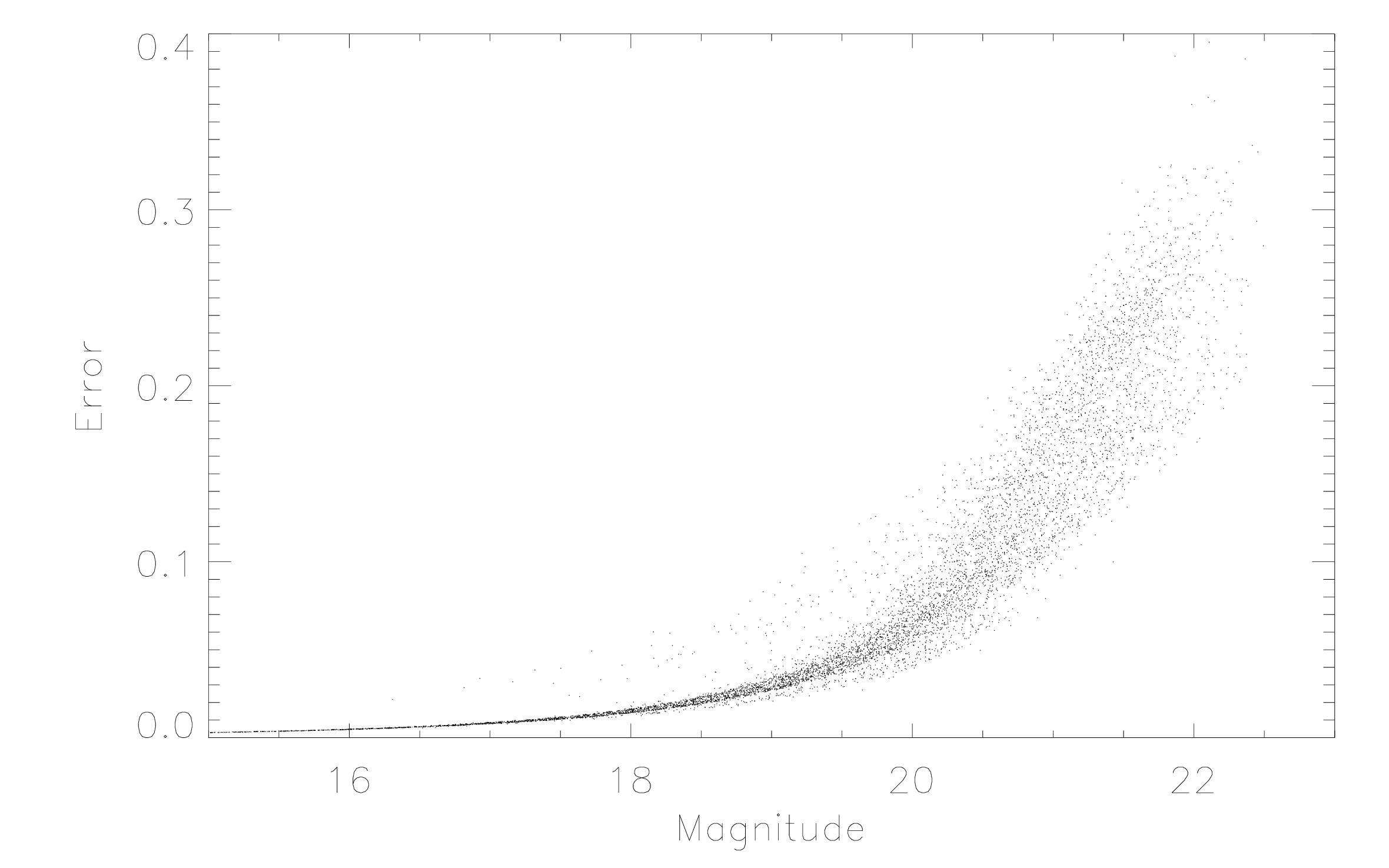} 
  \includegraphics[angle=0,scale=0.3]{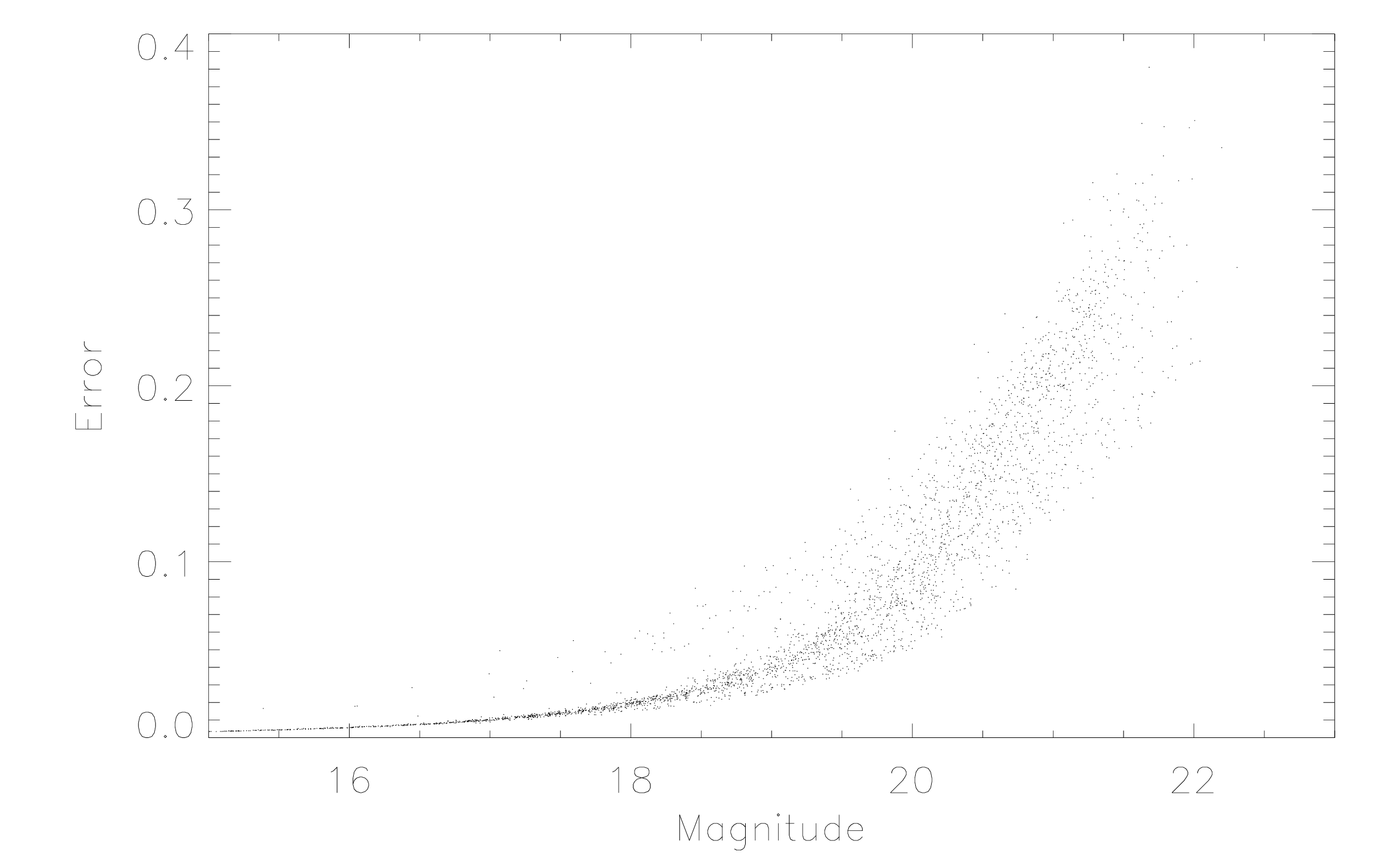}
  \caption{The photometry uncertainties (from poisson statistics)
    vs. magnitude in the SAGES project in the $u_s$-band (top) and
    $v_s$-band (bottom) for a typical field. The limiting magnitude is
    $\sim 17$ mag for both the 
    $u_s$-band and the $v_s$-band of SAGES, with uncertainties of 0.01
    mag, which corresponds to S/N$\sim100$.} 
  \label{fig13}
\end{figure}

Figure~\ref{fig13} is the plot of the photometry uncertainties versus
the magnitude of SAGES in the $u_s$-band and the $v_s$-band. Here
  the photometry uncertainties are just from the poisson statistics,
  which are not the rms of repeat observations, since for most sources
  we just observed for only one time. It can be seen  
that the limiting magnitude is $\sim 17$ mag for both the $u_s$-band 
and the $v_s$-band of SAGES, for the uncertainties of the magnitude 
of 0.01 mag, which correspond to the S/N$\sim100$ and different
  from the complete magnitude.

\subsection{Astrometry}
\label{astro.sec}

The astrometric calibration is realized in two steps. Firstly, the
pipeline will work out a linear solution assuming the image has no
distortion, based on the information stored in image headers,
including the telescope pointing coordinates, the rotation angle, and
the pixel scale. For further corrections, the pipeline employs the
package named SCAMP (Software for Calibrating AstroMetry and
Photometry, https://www.astromatic.net/software/scamp/), which is a
computer program that computes astrometric projection parameters.  
Otherwise, SCAMP is mature and robust software that is widely used in
astrometric calibration by matching SExtractor's output catalog with
an online or local reference catalog \citep{ber96}.

Our pipeline runs SCAMP twice: for the first run, we use a loose
criterion for cross-matching detected sources with reference sources,
which is in order to maintain enough stars in the image for
calibration; for the second run, we applied a more strict criterion to
obtain a more precise solution based on the previous results. Table~\ref{t2.tab}
shows the configuration parameters for SCAMP in our pipeline for the
two runs. It is found that DISTORT DEGREES $=3$ is the most proper
configuration after a series of tests \citep{zheng18,zheng19}. 

In our astrometric pipeline, we use the Position and Proper Motion
Extended (PPMX) \citep{roser} as the astrometric reference, which
contains $\sim$18 million stars. The reference stars are distributed
across the whole sky with the astrometric accuracy of
$\sim0^{\prime\prime}.02$ in both R.A. and Dec \citep{zheng18,zheng19}.  

Although Pan-STARRS DR1 (PS1) \citep{cham} has an accurate position
estimate, i.e., both uncertainties are lower than $0^{\prime
  \prime}$.005 in both R.A. and Dec directions, PS1 does not have
information on proper motions. Thus we do not use the PS1 catalog as the
astrometric reference, and we use it as the flux reference (see 
Section~\ref{fluxc.sec}).  We use PPMX as the astrometric
  reference. In fact, in the matching with our SAGES, $> 85$\% of the
  PPMX stars are in the magnitude range of 10.0$<$ V$<$ 15.0 mag. As
  the Gaia DR3 has been released, we may improve the precision of the
  astrometric reference in future work. Due to the lack of providing
  the astrometric residuals, we cannot check the solution is correct or not. In
  addition, besides SCAMP, we also developed another astrometric
  calibration method SIP (Simple Imaging Polynomial) \citep{shupe} for
  providing an astrometric solution, since SCAMP does not provide
    the fitting errors which can be used to judge the results are good
    enough or not. Therefore, in the astrometric pipeline, we applied
  both methods to make sure the solution is correct and accurate.

In our astrometric pipeline, we first convert the original
pixel coordinates (x, y) with origin at the left-bottom corner of one
single frame (one chip), to intermediate pixel coordinates ($u_s$,
$v_s$) with the origin at the center of the whole FOV. The second step
is to convert the ($u_s$, $v_s$) to the intermediate world coordinates
($\epsilon$, 
$\eta$) with parameters CD$_{ij}$ recorded in the fits header. The
  parameters CD present  a matrix which transfers from (u, v) to
  ($\epsilon$, $\eta$). Initially, they can be computed by the rotate
  angle and the pixel-scale of detector. They are provided in headers
  after the header fixing process.  Finally,
the intermediate world coordinates ($\epsilon$, $\eta$) are projected to the world
coordinates ($\alpha$, $\delta$), with the projection type of
``TAN''. In order to resolve the non-linear correlation transformation
between ($\epsilon$, $\eta$) and ($\alpha$, $\delta$), we applied the SIP
convention to represent image distortion as it introduces high-order 
correction polynomials f and g to $u_s$, $v_s$ to express the distortion, 
as in formula 1 \citep{zheng18,zheng19}. 

$$      
\left(                 
  \begin{array}{c}   
    \epsilon \\  
    \eta \\  
  \end{array}
\right)                 
=CD\times
\left(                 
  \begin{array}{c}   
    u+f(u,v)\\  
    v+g(u,v)\\  
  \end{array}
\right)                 
$$	

In our transform of the astrometry, $A_{pq}$ and $B_{pq}$ are used as the
coefficients of $u^pv^q$ as shown in Formula 2 to determine polynomials f
and g, in which $N_A$ and $N_B$ are the highest order to correct $u_s$ and
$v_s$. We adopt appropriate parameters NA $=$ NB $= 3$ for the SAGES. 

$$ f(u,v)=\sum _{p,q} A_{pq} \cdot u^pv^q, 2 \le p+q \le N_A $$
$$ g(u,v)=\sum _{p,q} B_{pq} \cdot u^pv^q, 2\le  p+q \le N_B $$

\begin{figure}
  \centering
  \includegraphics[angle=0,scale=0.25]{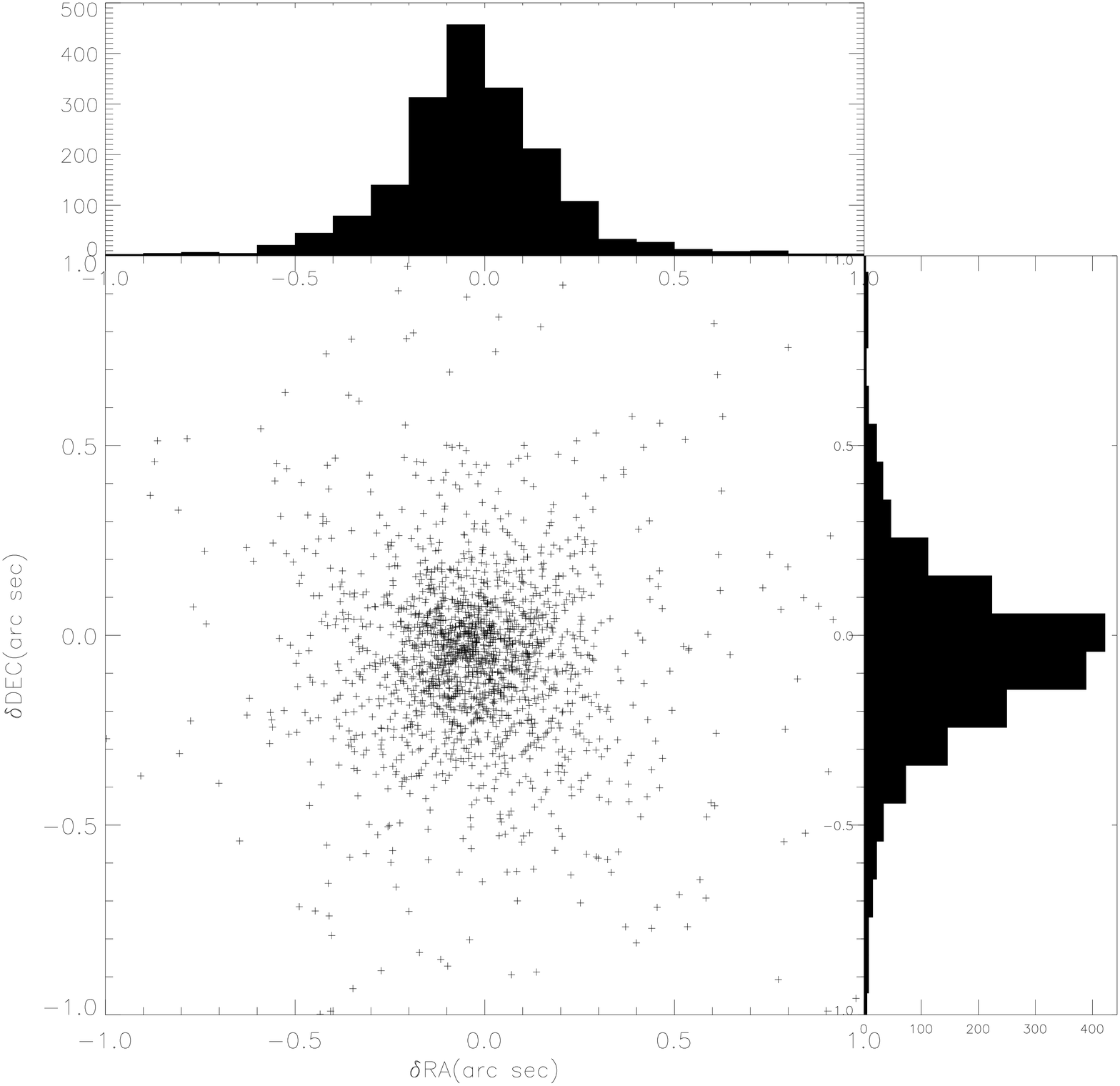}
  \includegraphics[angle=0,scale=0.25]{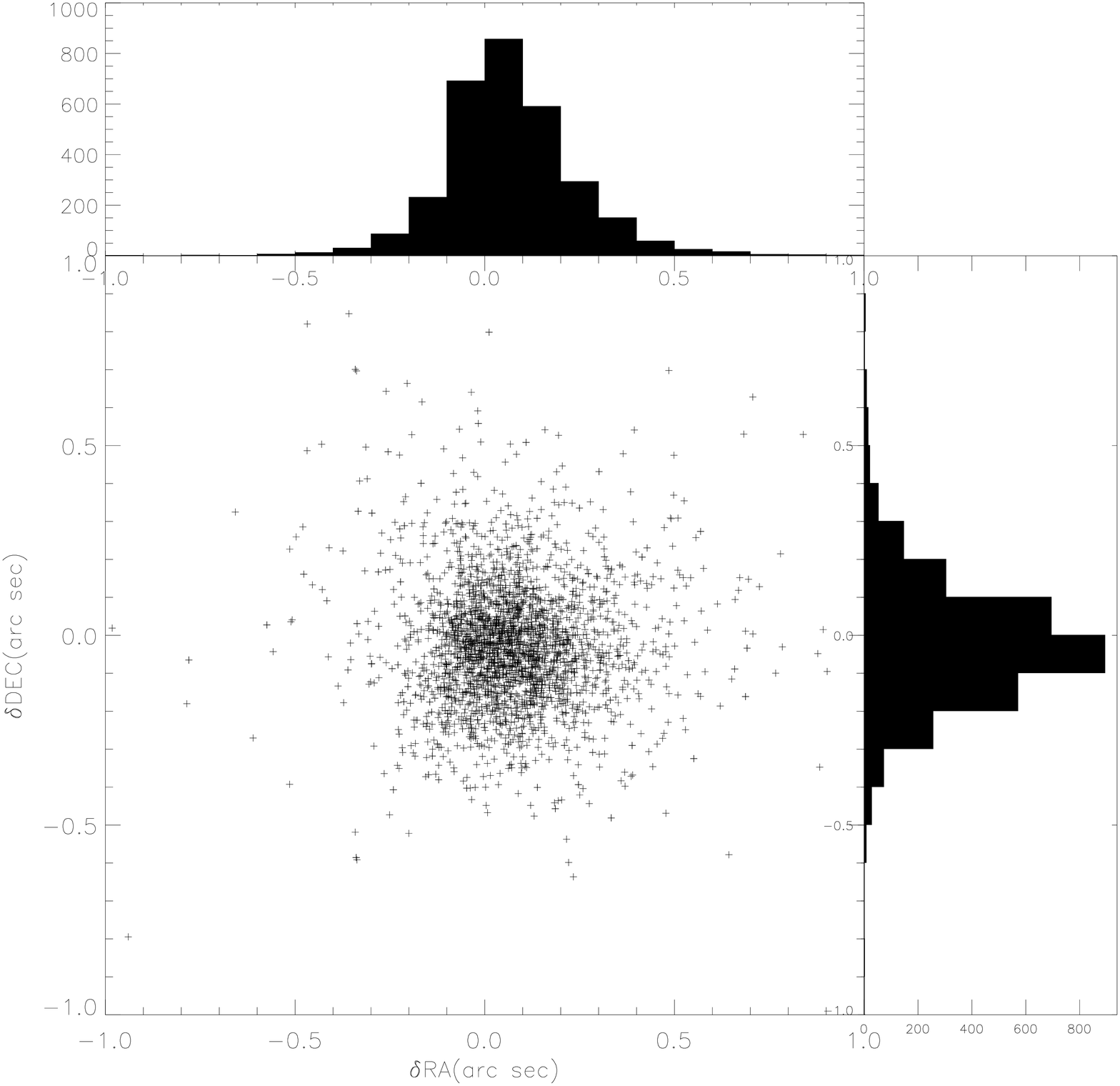}
  \caption{The precision of the astrometry in the SAGES project in the
    $u_s$-band (top) and $v_s$-band (bottom) for one typical
    image. The external astrometric errors are estimated by
      comparing the difference between the coordinates from our
      catalog and those from reference catalog PPMX.} 
  \label{fig10}
\end{figure}

We use the multiple visits, no matter if in the same
band or not, to estimate the differences in coordinates between different visits.
Figure~\ref{fig10} shows a typical internal astrometric error of one image,
in which it shows that $\Delta R.A. = 0^{\prime\prime}.014 \pm
0^{\prime\prime}.145$ and $\Delta Dec =-0^{\prime\prime}.002 \pm
0^{\prime\prime}.166$. The external astrometric errors are 
estimated by comparing the difference between the coordinates from
our catalog and those from reference catalog PPMX. Figure~\ref{fig12} shows
a typical distribution of external astrometric calibration
errors in one observing field. It can be seen that the standard
deviations is quite small, i.e., $\sim0^{\prime\prime}.1$ in both
R.A. and Dec directions, as marked in the lower-left panel, which includes
both internal and external astrometric uncertainties in both
directions \citep{zheng18,zheng19}. In the future work, it can be improved 
when employing Gaia DR3 as a reference catalog.

\begin{figure}
  \centering
  \includegraphics[angle=0,scale=0.45]{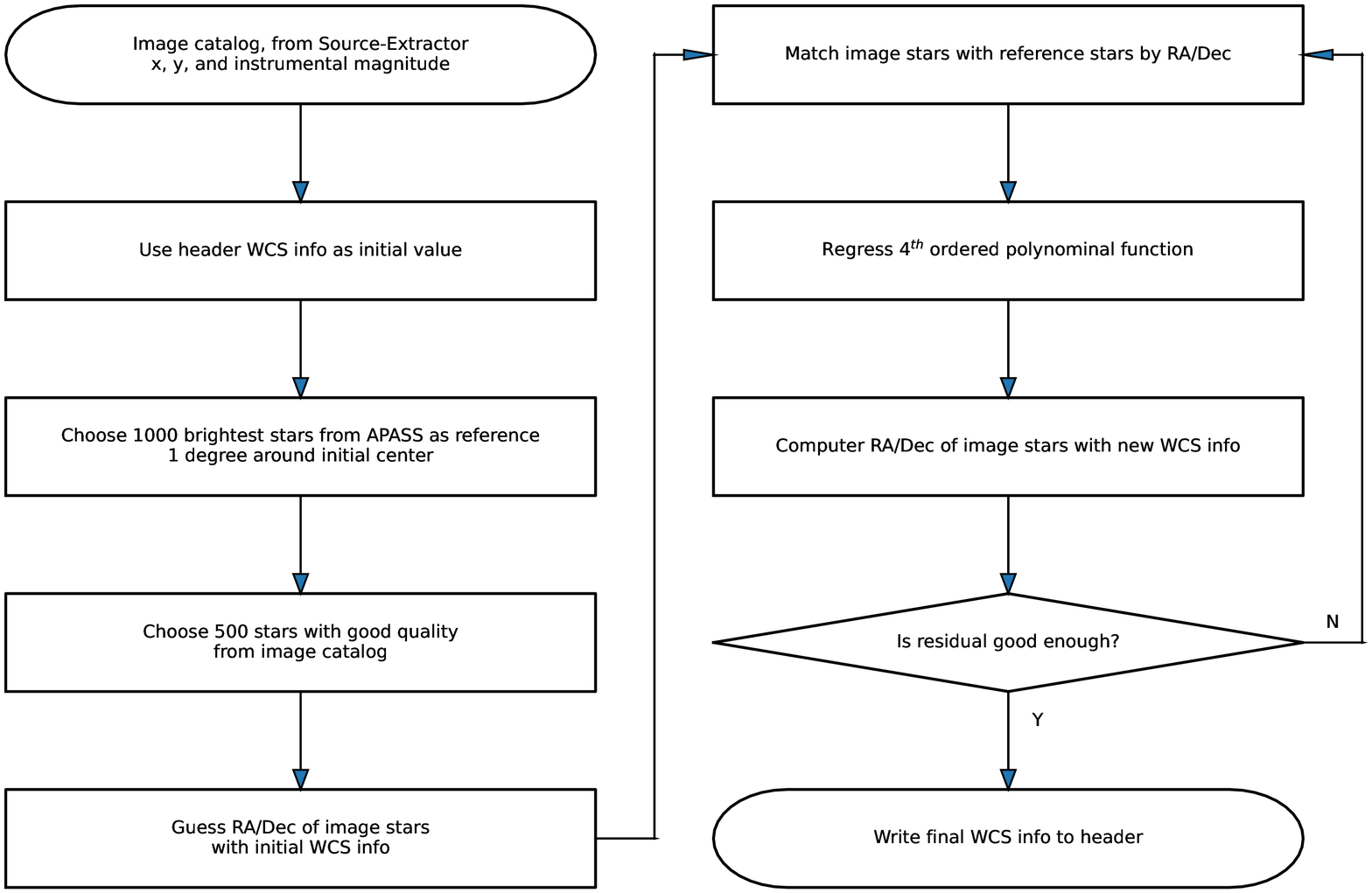}
  \caption{The flow chart of the SIP for SAGES astrometry 
    after two runs of the SCAMP. In our work, we applied the pipeline
  based on the SCAMP and our routine SIP.}
  \label{fig11}
\end{figure}

Figure~\ref{fig11} is the flow chart of the SIP for the SAGES astrometry, 
which adopt the Source Extractor results as the input catalog and applies
the fits header as the WCS initial value. The task SCAMP is applied
twice and we use the PPMX as the reference catalog. 

Table~\ref{t2.tab} shows the configuration parameter of astrometry
table for SCAMP, which has been applied for our data reduction pipeline. 

\pagestyle{empty}
\begin{deluxetable}{llll}
  \rotate
  \tablecolumns{4} \tablewidth{0pc} \tablecaption{The configuration
    parameter of astrometry table for SCAMP applied in our pipeline.
    \label{t2.tab}}
  \tablehead{
    \colhead{Keywords} &  \colhead{Round 1} &   \colhead{Round 2} & \colhead{Notes}}
  \startdata
MATCH & Y & Y & Module match or not \\
MATCH\_NMAX & 0 & 0 & up bound of cross match \\
PIXSCALE\_MAXERR & 2 & 1.5 & Max error of pixel scale \\		  
POSANGLE\_MAXERR & 5.0 & 2.0 & Max error of position angle (degree) \\
POSITION\_MAXERR & 10.0 & 1.0 & Max error position \\
MATCH\_RESOL & 0 & 0 & matching resolving  \\
MATCH\_FLIPPED & N & N & Allow axis flipping in match or not  \\
CROSSID\_RADIUS & 25.0 & 25.0 & Cross identification radius   \\
SOLVE\_ASTROM & Y & Y & Solve astrometric solution or not     \\
PROJECTION\_TYPE & SAME & SAME & Projection type             \\
DISTORT\_DEGREES & 3 & 3 & Degree of Distortion Polynomial   \\
  \enddata
\end{deluxetable}

\subsection{Flux Calibration}
\label{fluxc.sec}

In this work, using the spectroscopic data from the Large Sky Area
Multi-Object Fiber Spectroscopic Telescope (LAMOST;
\citealt{cui12,zhao12}) DR5, photometric data from  
the Gaia DR2 (\citealt{2016A&A...595A...1G, 2018A&A...616A...1G}), and
over-lapping observations of the SAGES, we first perform the relative
flux calibration of SAGES  
DR1  $u_s$/$v_s$ bands by combining  the Stellar Color Regression (SCR; 
\citealt{yuan15}) method and the Ubercalibration method
\citep{pad08}. The absolute calibration is then carried out by
comparing with synthetic colors from the MILES library
(http://research.iac.es/proyecto/miles/) \citep{SB06}.  

 In fact, \cite{yuan15} have proposed the spectroscopy-based SCR
  method to perform precise color calibrations by using millions of
  spectroscopically observed stars as color standards, with the
  star-pair technique \citep{yuan13}, which considers that stellar
  colors can be accurately predicted by large-scale spectroscopic
  surveys, e.g., SDSS, LAMOST.

Combining with the accurate and homogeneous photometric data from {\it
  Gaia} DR2 and Early Data Release 3 (EDR3;
\citealt{2021A&A...649A...1G,2021A&A...650C...3G}), the method can
further  accurately predict the magnitudes of stars in various
passbands and perform  high precision photometric calibration. Such
method has been applied to a number of surveys, including the SDSS
Stripe 82 \citep{yuan15, Huang22}, the SkyMapper Southern
Survey DR2 \citep{2021ApJ...907...68H}, the Gaia DR2 and EDR3  
(\citealt{2021ApJ...909...48N,2021ApJ...908L..14N}), and the PS1 DR1
\citep{Xiao and Yuan(2022)}. A precision of 1 to a few mmag is usually
achieved. A recent review of the method and its implementations can be  
found in \cite{2022arXiv220601007H}.  

The Ubercalibration method  was originally developed for SDSS, and
achieved a precision of 1\% in the  
griz bands, and 2\% in the $u$ band \citep{pad08}. 
It requires a significant amount of over-lapping/repeating observations and 
assumes that the physical magnitude of the same object under different
observational conditions should be the same.  
\cite{Schlafly(2012)} have applied the method to PS1 catalogs,
achieving a precision of better than 1\%. The method has also been
applied to the Beijing–Arizona Sky Survey (BASS)
\citep{2017PASP...129...064101, ZhouZhimin(2018)}, achieving a
precision of better than 1\%. A detailed summary and discussion of
limitations of the method can be found in \cite{2022arXiv220601007H}. 

We combine the two aforementioned methods to perform the relative flux
calibration of SAGES DR1  $u_s$/$v_s$ bands. A detailed description of
the calibration process will be presented in a separate paper (Yuan
H. et al. in preparation). Here we briefly outline the calibration
strategy. 

The calibration process of SAGES DR1 is carried out for each gate
separately in the  $u_s$ and $v_s$ bands. 
We assume that the relative calibrated magnitude of an object $m_{\rm rel}$ 
can be derived from its instrumental magnitude $m$ by 
  \begin{eqnarray}
  m_{\rm rel}=m+zp\_frame(i)+zp\_gate\_corr(i,j)+flat\_corr(day,j,X,Y),  \label{cali_mod}
  \end{eqnarray}
where $zp\_frame(i)$ is the zero point of the $i$-th frame, $zp\_gate\_corr(i,j)$ 
is the zero point correction of the $j$-th gate of the $i$-th frame, and 
$flat\_corr(day,j,X,Y)$ is the daily star flat correction of the
$j$-th gate and depends on CCD position (X, Y).   
Taking u$_s$ band for example, the detailed calibration strategy is
listed as below:  
\begin{enumerate}
\item[a.] Combine the SAGES DR1 photometric data with LAMOST DR5 and
  the Gaia DR2 to create a sample for the flux-calibration, which
  have data from all the three database above. Select dwarfs as the
calibration stars with the following constraints: (1)  
$5300<T_{\rm eff}<6800$\,K and $-0.8<\rm [Fe/H]<0.2$, a narrow
temperature and  
metallicity range for robust fitting of stellar colors as a function
of stellar atmospheric parameters; and (2) Signal-to-noise ratio ($\rm
S/N_{g}$) of the LAMOST spectra $>20$. About 1.5 and 1.1 million
calibration stars are selected for the $u_s$ and $v_s$ bands, respectively. 

\item[b.] Fit stellar intrinsic color -- stellar atmospheric parameters relation 
$(G_{\rm BP}-u_s)_0$ = $f(T_{\rm eff}, \rm [Fe/H])$ using stars from two-well 
selected neighboring fields which have more LAMOST targets for
  calibration. Here $f(T_{\rm eff}, \rm [Fe/H])$ is a second-order  
two-dimensional polynomial with six free parameters. The
  neighboring fields means that they have similar position and
  observing time, thus they should have similar zero-points so that
  they can  be regarded as one field and have more sources.
  \begin{eqnarray}
  f(T_{\rm eff}, \rm [Fe/H])=&{a_0}\cdot T_{\rm eff}^2+{a_1}\cdot {\rm [Fe/H]}^2+{a_2}\cdot T_{\rm eff}\cdot {\rm [Fe/H]}+{a_3}\cdot T_{\rm eff}+{a_4}\cdot {\rm [Fe/H]}+{a_5}.  \label{color_mod}
  \end{eqnarray}
  
\item[c.] Estimate the predicted magnitudes for all the calibration
  stars with a typical precision of $2-3\%$, and obtain zero point of
  each gate of each image file. 

\item[d.] Construct the daily star flat for each gate
  $flat\_corr(day,j,X,Y)$ using second-order two-dimensional
  polynomials as a function of $X$ and $Y$.   

\item[e.] After star flat correction, obtain zero point of each image file 
$zp\_frame(i)$ and zero point correction $zp\_gate\_corr(i,j)$ of each gate 
and each image file. 

\item[f.] Go to step b, reconstruct the relation, and iterate. 

\item[g.] Combine the ubercalibration method and the SCR method to
  further improve the calibration, particularly for the image files
  with nil or a small number of LAMOST standard stars.  

\item[h.] At last, perform absolute calibration by comparing the color-color 
relationship ($G_{\rm BP}-u_s~vs.~G_{\rm BP}-G_{\rm RP}$) between the
observed and synthesized cases from the empirical MILES spectral
library. 
\end{enumerate}

Reddening correction in steps b and c is performed using $E(BP-RP)$
and empirical temperature- and reddening- dependent reddening
coefficients, both of which are obtained using the star-pair technique
similar to the one in \cite{sun2022}.   
Applying the above procedure, we have achieved an internal calibration
precision of around 5 mmag for the SAGES DR1 data by comparing
repeated observations. 

\begin{figure}
  \centering
 \includegraphics[angle=0,scale=0.88]{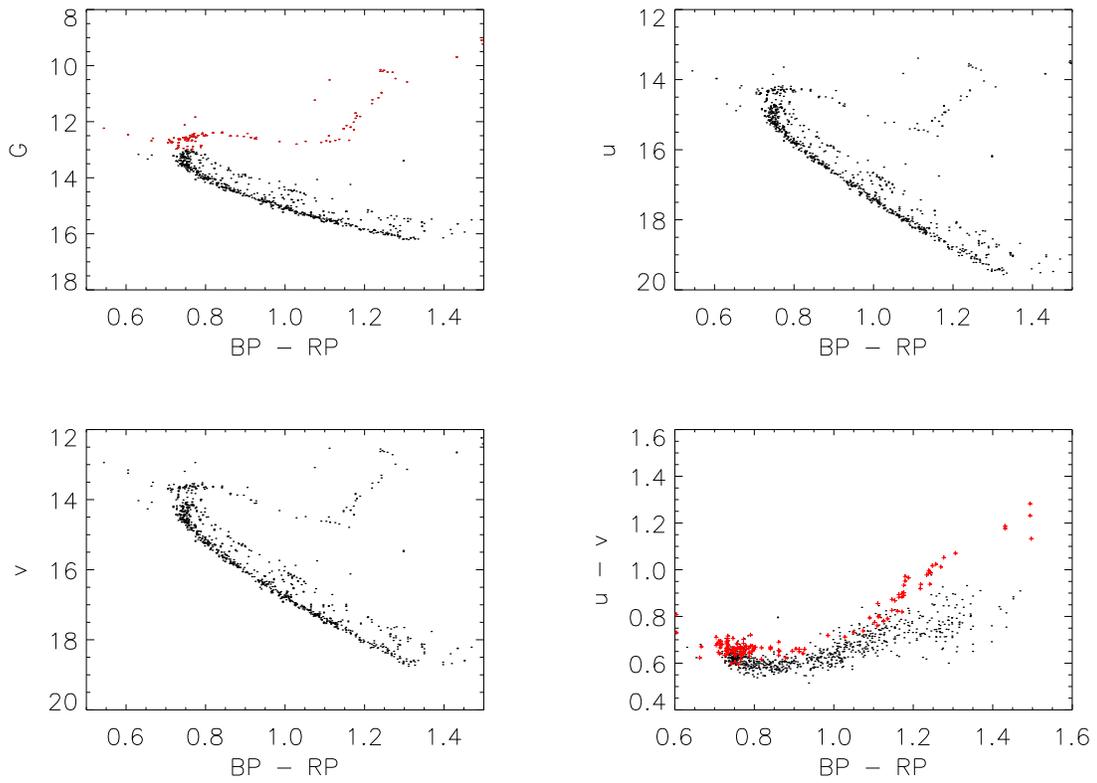}
  \caption{The color-magnitude of the M67 cluster with SAGES
    and Gaia DR2 photometry. The red dots are the giants and
    the black dots are the dwarfs. The giants and
    dwarfs can be well separated in the color-color diagram (bottom
    right).}
  \label{fig14}
\end{figure}

Figure~\ref{fig14} shows the color-magnitude and color-color diagrams 
of M67 with the SAGES and Gaia DR2 photometry but the DR3
  will not change much. It can be seen that our
SAGES photometric precision is comparable to that of the Gaia
BP/RP. The red dots are the giants and black dots are the dwarfs. We
can see that the giants and dwarfs can be separated very well. 

\subsection{Cross-Matched External Catalogs}
\label{cro.sec}

As in SAGES DR1, we mainly cross-match the master table with PS1
catalogs as its observing depth is comparable to our $u_s$ and $v_s$
passband. Although we have the complete $gri$-passband observations of
the 1-m telescope at the Nanshan station of XAO, it is not deep
enough. Previously we planned to combine our observations of XAO to
complement the SDSS imaging survey. However, it is well known that
the combination of the two kinds of survey data may not be homogeneous
for the observations depth. Also we need the transformations of the
passband between the Nanshan data and the SDSS data. 

Thus we combined our $u_s$ and $v_s$ band data with the PS1, which is
homogeneous and sky coverage is large enough, i.e, $\sim3/4$ of the total 
sky. For large photometric catalogs, we determine the
matching external object and record its ID and projected distance
within the master table. In our work, we match in the reverse direction
and distance in the external catalog as DR1-ID. However, the
  proper motion information is not included in the catalog. The
maximum distance for all cross-matches is $2^{\prime\prime}$, motivated 
by the region in which our 1D PSF magnitudes may be affected.

Such matching process has resulted in a catalog with half of the
original sources. There are 39,857,439 sources and 42,457,643 sources
left in the $u_s$ and $v_s$ passbands respectively after combining
with the PS1 catalog.  
 
\section{Data Products of SAGES}
\label{dat.sec}

\subsection{The Contents of SAGES DR1}
The SAGES Data Release 1 contains the total photometry from a total
of 36,092 image exposures, including 23,980 images in $u_s$ passband
and 17,565 images in the $v_s$ passband. In total we have 41,545
frames for $u_s$ and $v_s$ bands. Among these images, 36,092 frames
have high quality and involve the flux calibration.
After combining with the PS1 catalog, the SAGES Master catalog thus contains
a total number of 54,215,563 sources which will be actually
  released in this paper. 

\subsubsection{The SAGES Master Table}
The SAGES DR1 total photometric table contains about one hundred million 
unique astrophysical objects in either SAGES $u_s$ and $v_s$
bands. The astrometric calibration of DR1 is 
accomplished and the median offset between our positions and those in Gaia DR2 is
0$^{\prime\prime}$.16 for  all objects and 0$^{\prime\prime}$.12 for
bright, well-measured objects. In the next release, we plan to switch
the astrometric reference frame to Gaia DR3. 

Table~\ref{t3.tab} shows the header of the SAGES master catalog, including the 
parameters of NUMBER, RA, RA$\_$ERR, DEC, DEC$\_$ERR, U$\_$COUNT,
U$\_$MAG$\_$AUTO, U$\_$ERR$\_$ISOCOR, U$\_$ERR$\_$ISOCOR, etc. as
well as the information of the PS1 catalog, such as the photometry, errors and
IDs.

\pagestyle{empty}
\begin{deluxetable}{ll}
  \rotate
  \tablecolumns{2} \tablewidth{0pc} \tablecaption{The contents and
    descriptions of the SAGES Master catalog.
    \label{t3.tab}}
  \tablehead{
    \colhead{Parameters} & \colhead{Descriptions of SAGES Master Catalog}}
  \startdata
NUMBER & Running object number \\
RA & Right ascension of object (J2000) \\
$\rm RA\_ERR$ & the Error of Right ascension \\
DEC & Declination of the object (J2000) \\
$\rm DEC\_ERR$ & the Error of Declination \\
$\rm u\_FLUX$ & data counts of $\rm AUTO\_FLUX$ \\
$\rm u\_MAG\_AUTO$ & AUTO magnitude in the $u_S$ passband \\
$\rm u\_ERR\_AUTO$ & photometry error of AUTO magnitude \\
$\rm u\_MAG\_ISOCOR$ & Isophotal magnitude in $u_S$-band \\
$\rm u\_ERR\_ISOCOR$ & photometry error of Isophotal magnitude in $u_S$-band \\
$\rm u\_MAG\_APERCOR$ & aperture magnitude in $u_S$-band  \\
$\rm u\_ERR\_APERCOR$ & photometry error of aperture magnitude in $u_S$-band \\
$\rm u\_MAG\_PETRO$ & Petrosian-like elliptical aperture magnitude in $u_S$-band  \\
$\rm u\_ERR\_PETRO$ & photometry error of Petrosian-like elliptical aperture magnitude in $u_S$-band  \\
$\rm u\_FLAGS$ & Combination method for flags on the same object: 17 – arithmetical OR,\\ 
 & 18 – arithmetical AND, 19 – minimum of all flag values, 20 – maximum of all flag values,\\ 
 & 21 – most common flag value  \\
$\rm v\_FLUX$ & data counts of $\rm AUTO\_FLUX$     \\
$\rm v\_MAG\_AUTO$ & AUTO magnitude in the $v_S$ passband \\
$\rm v\_ERR\_AUTO$ & photometry error of AUTO magnitude \\
$\rm v\_MAG\_ISOCOR$ & Isophotal magnitude in $v_S$-band \\
$\rm v\_ERR\_ISOCOR$ & photometry error of Isophotal magnitude in $v_S$-band  \\
$\rm v\_MAG\_APERCOR$ & aperture magnitude in $v_S$-band   \\
$\rm v\_ERR\_APERCOR$ & photometry error of aperture magnitude in $v_S$-band  \\
$\rm v\_MAG\_PETRO$ & Petrosian-like elliptical aperture magnitude in $v_S$-band  \\
$\rm v\_ERR\_PETRO$ & photometry error of Petrosian-like elliptical aperture magnitude in $v_S$-band \\
$\rm v\_FLAGS$ & Combination method for flags on the same object: 17 – arithmetical OR,\\ 
 & 18 – arithmetical AND, 19 – minimum of all flag values, 20 – maximum of all flag values, \\
 & 21 – most common flag value  \\
$\rm g\_MAG\_PS1$ & g mag of Pan-STARRS    \\
$\rm g\_ERR\_PS1$ & photometry error of g mag of Pan-STARRS catalog \\
$\rm r\_MAG\_PS1$ & r mag of Pan-STARRS    \\
$\rm r\_MAG\_PS1$ & photometry error of r mag of Pan-STARRS catalog \\
$\rm i\_MAG\_PS1$ & i mag of Pan-STARRS    \\
$\rm i\_ERR\_PS1$ & photometry error of i mag of Pan-STARRS catalog \\
$\rm ID\_PS1$ & ID from Pan-STARRS catalog  \\

  \enddata
\end{deluxetable}

The photometric measurement can be qualified through the SE FLAGS
(reliable 0 to 3, and caution should be taken for $>4$) .

\pagestyle{empty}
\begin{deluxetable}{ll}
  \rotate
  \tablecolumns{2} \tablewidth{0pc} \tablecaption{The contents and
    descriptions of the SAGES merge Bits table.
    \label{t4.tab}}
  \tablehead{
    \colhead{Bits} & \colhead{Represents Meaning}}
  \startdata
0 & Detected only once     \\
1 & Has other close objects but rejected while merging     \\
2 & Flux not good enough, at least 1 source is out of 3$\sigma$    \\
3 & Position not good enough, at least 1 source is out of 3$\sigma$  \\
4-27 & Reserved   \\
28   & No matched object  \\
29-31 & Reserved \\
  \enddata
\end{deluxetable}

\subsubsection{Limitation of auto- and aperture magnitudes}

\begin{figure}
 \centering
  \includegraphics[angle=0,width=0.9\textwidth]{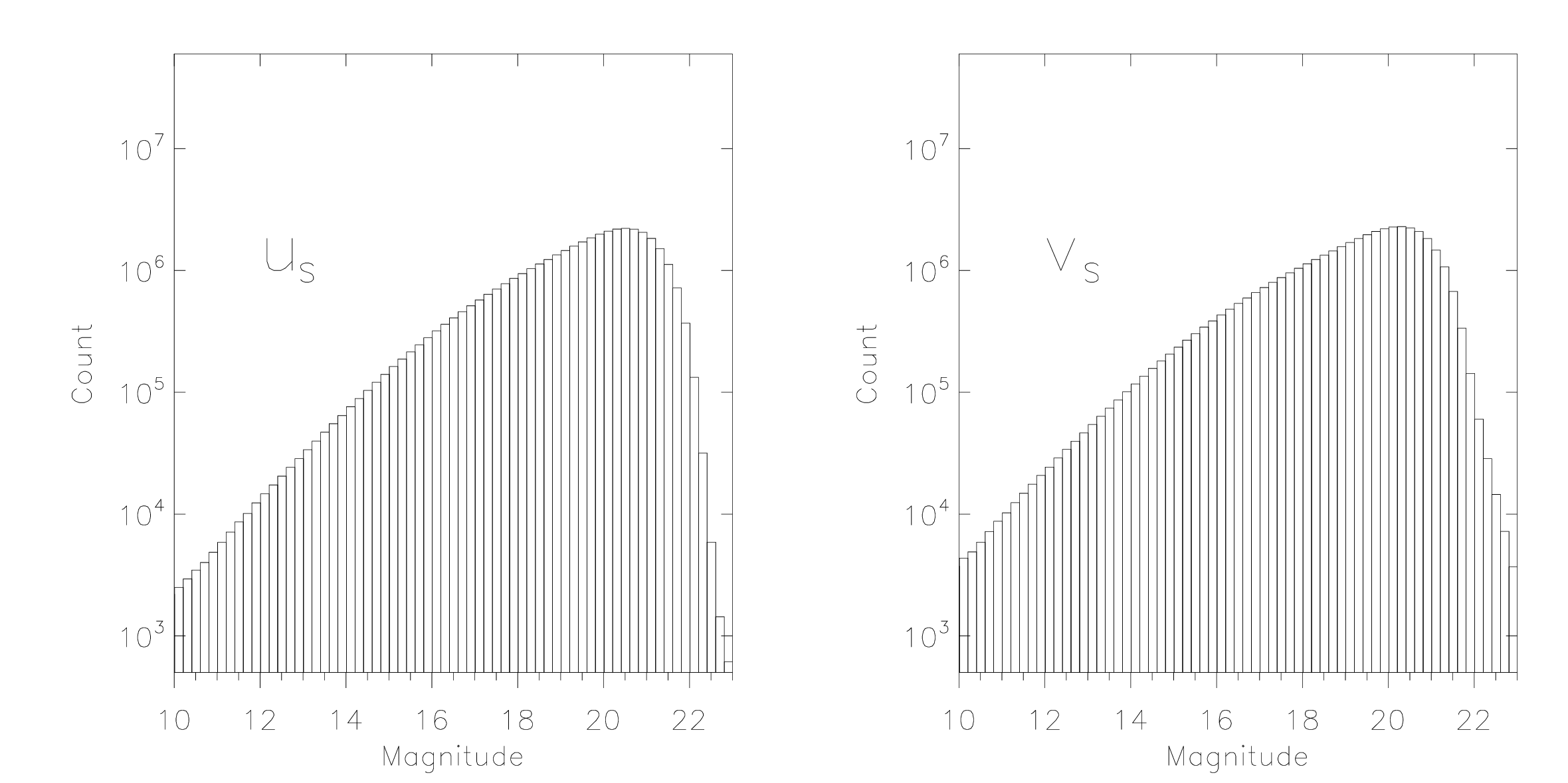}
  \caption{The magnitude distribution of the SAGES $u_s$ and $v_s$ bands 
  for all the sources of the SAGES catalog in all observed fields,
  even including the sources fainter than the "limiting magnitude". 
  The y-axis represents the counts for the photometry magnitude number. 
  It can be seen that the peaks are $u_s\sim21$ mag and $v_s\sim20.5$ mag 
  for our SAGES photometry.}
  \label{fig15}
\end{figure}

Figure~\ref{fig15} shows the magnitude distribution of the SAGES  $u_s$
and $v_s$ bands for all the sources of the SAGES DR1 catalog in all our
observing fields, including the sources fainter than the
  "limiting magnitude". We can see that for the $u_s$-band the limiting
magnitude is around 20 mag while for the $v_s$-band, the limiting
magnitude is around 21 mag.  
 
\begin{figure}
  \centering
  \includegraphics[angle=0,width=0.85\textwidth]{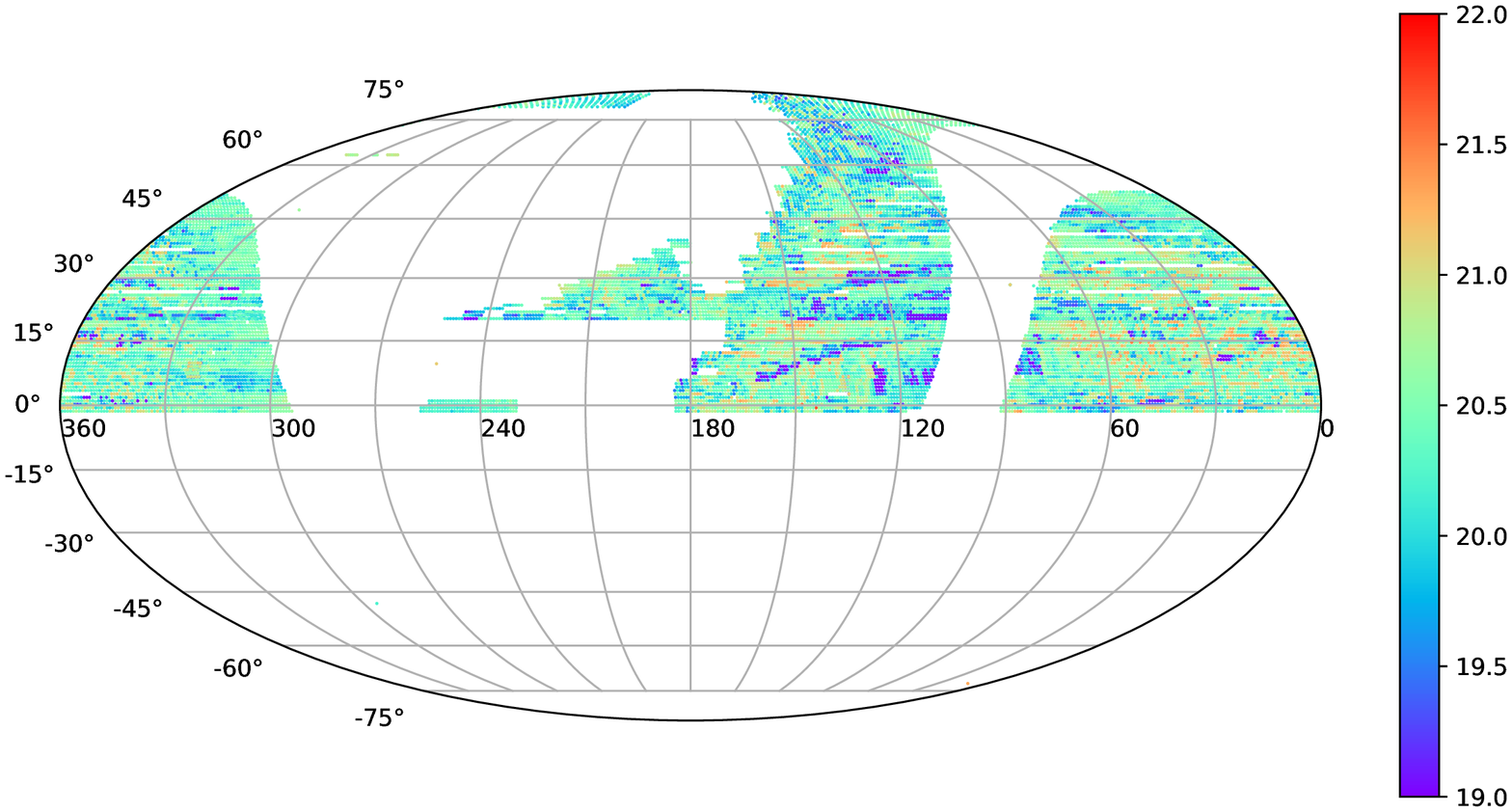}
  \caption{The observing area distribution of limiting magnitude in the
    SAGES $u_s$ passband for all fields for S/N=100, corresponding to
    the photometry error of 0.01 mag.} 
  \label{fig16}
\end{figure}

\begin{figure}
  \centering
 \includegraphics[angle=0,width=0.85\textwidth]{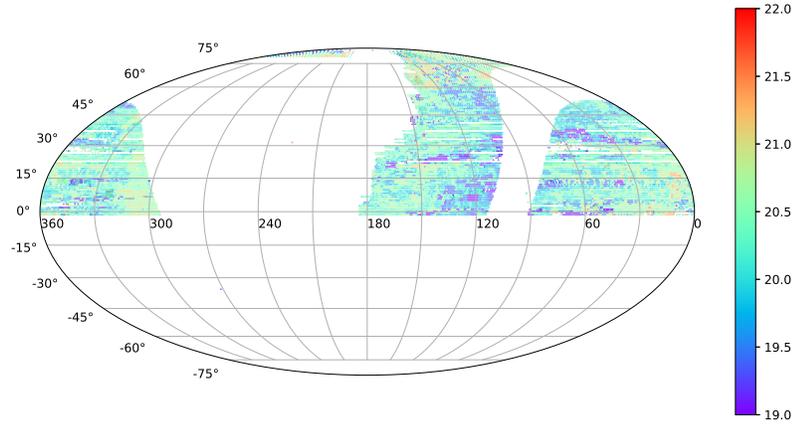}
  \caption{Similar to Figure~\ref{fig16} but for the SAGES $v_s$
    passband, limiting magnitude for S/N=100, corresponding to
    the photometry error of 0.01 mag.}
  \label{fig17}
\end{figure}

\begin{figure}
  \centering
  \includegraphics[angle=0,width=0.85\textwidth]{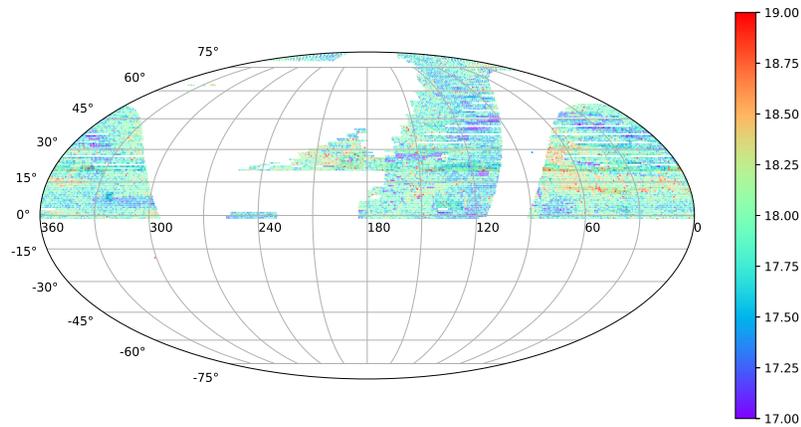}
  \caption{Similar to Figure~\ref{fig16} but for complete
    magnitude (which defines as the turnover in the magnitude
  histogram) of the SAGES $u_s$ passband, which is $\sim2.4$ mag deeper
    than that in Figure~\ref{fig16}.}
  \label{fig18}
\end{figure}

\begin{figure}
  \centering
  \includegraphics[angle=0,width=0.85\textwidth]{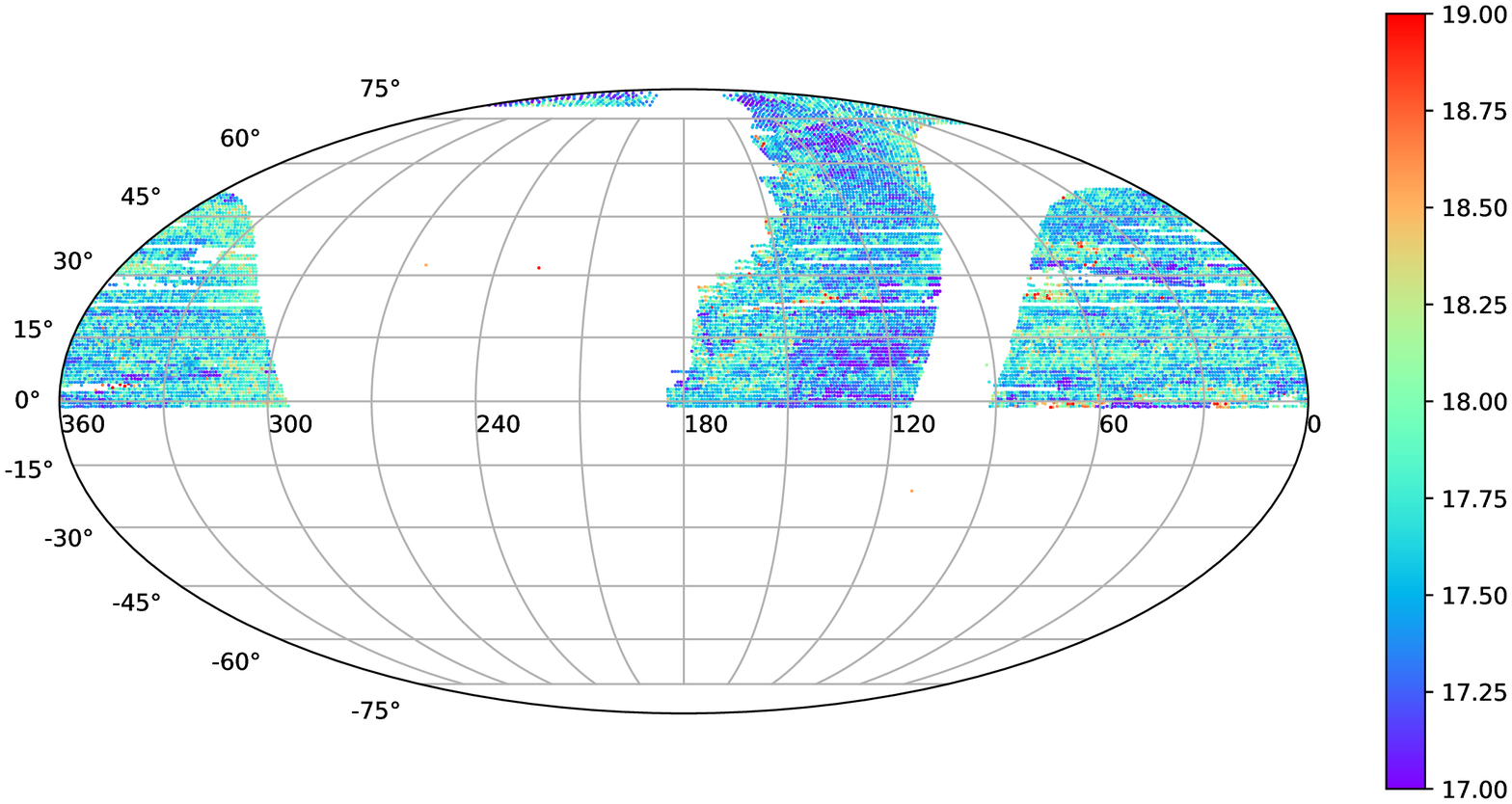}
  \caption{Similar to Figure~\ref{fig16} but for complete
    magnitude of the SAGES v$_s$ passband, which is $\sim2.4$ mag deeper
    than that in Figure~\ref{fig16}.} 
  \label{fig19}
\end{figure}

Figures~\ref{fig16}-\ref{fig19} are the distributions of observing depth 
(i.e. limiting magnitude of different S/N and complete magnitude,
(which defines as the turnover in the magnitude histogram)) of the
SAGES survey for the best magnitude for the 
$u_s$-/$v_s$-passbands.   

\begin{figure}
  \centering
  \includegraphics[angle=0,scale=0.5]{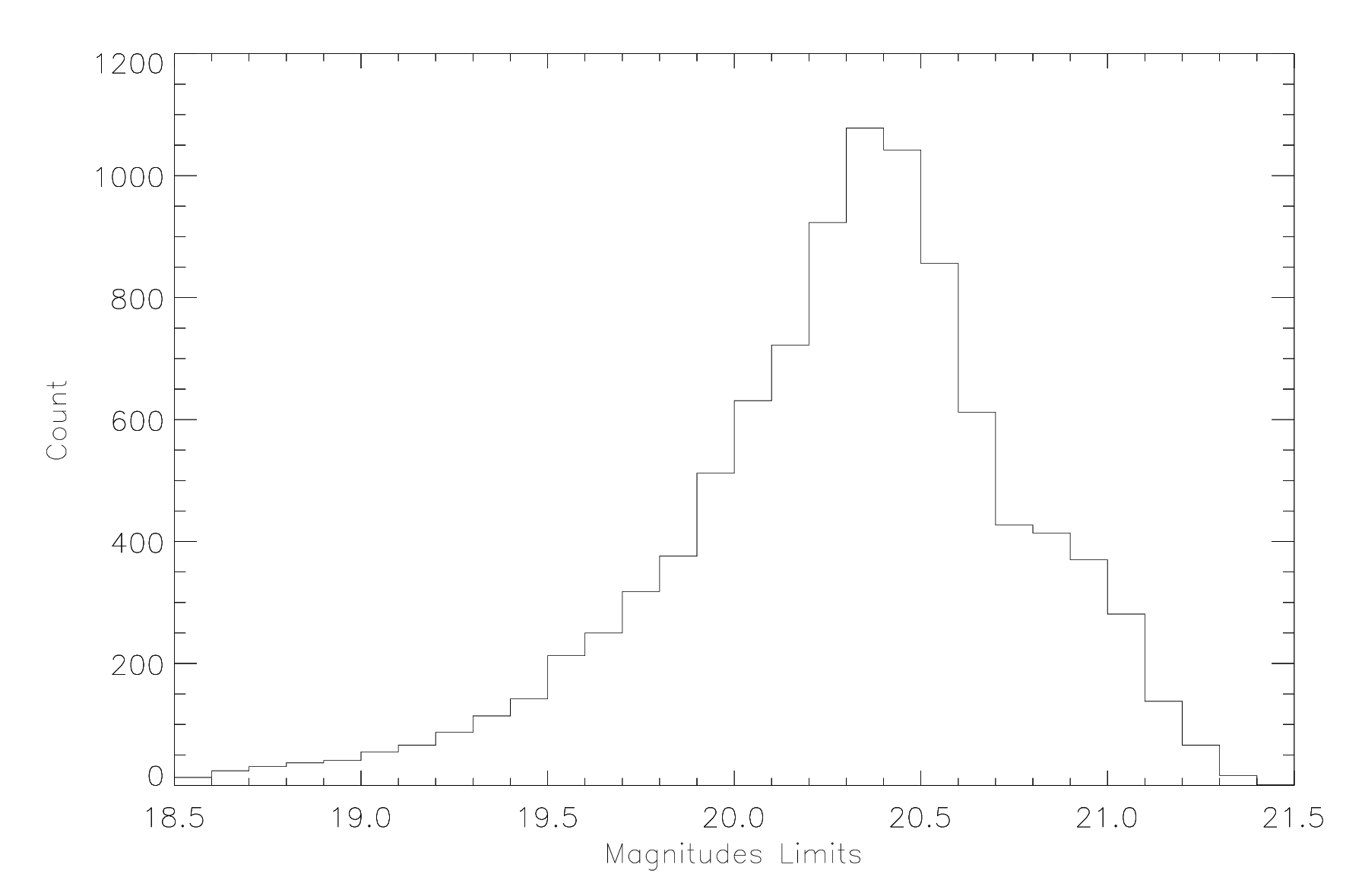}
  \caption{The distribution of the complete magnitude (which
      defines as the turnover in the magnitude histogram) of the SAGES $u_s$
    passband for all fields, with the median value of $u_s=20.4$ mag.} 
  \label{fig20}
\end{figure}

\begin{figure}
  \centering
  \includegraphics[angle=0,scale=0.5]{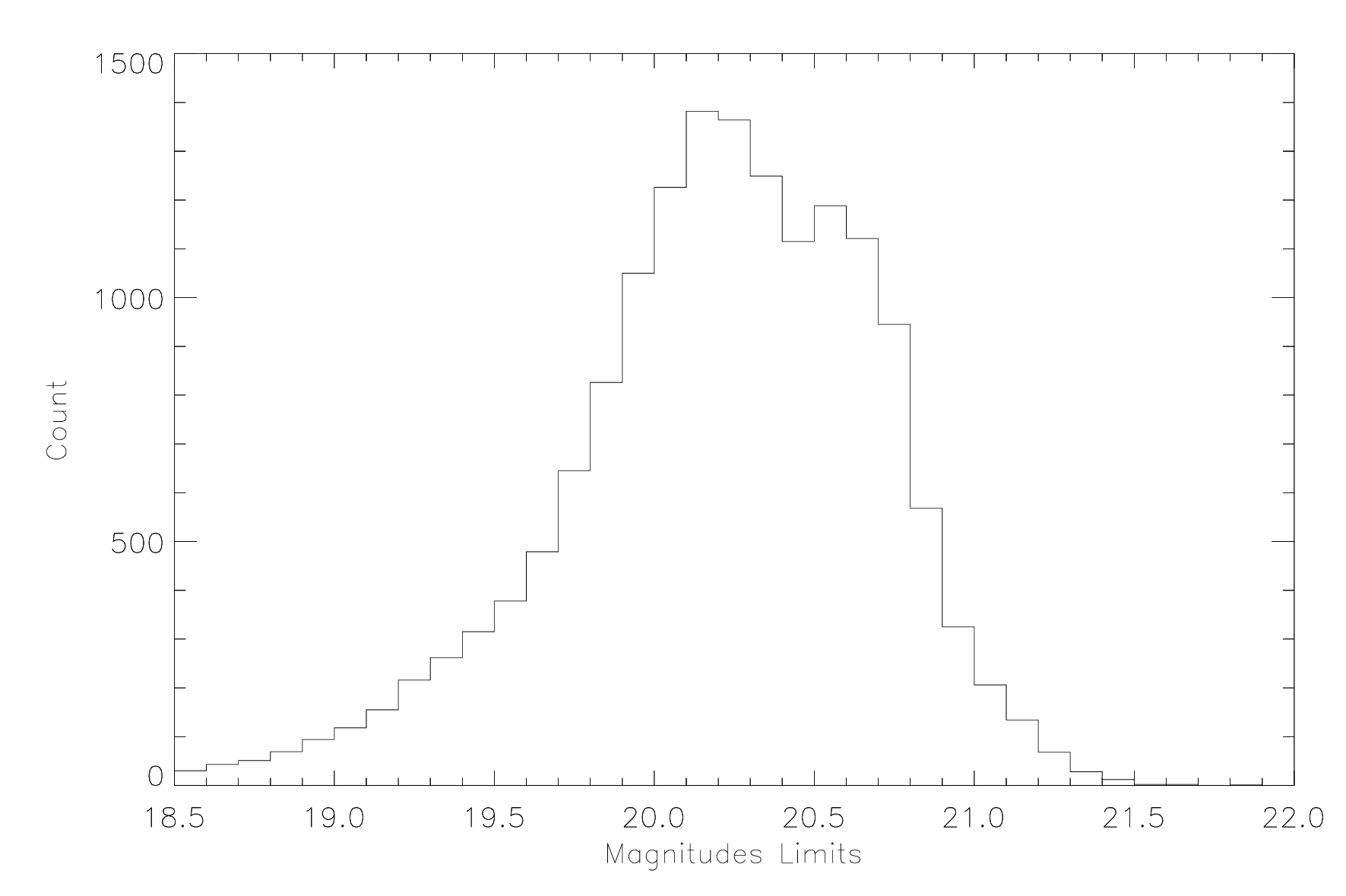}
  \caption{Similar to Figure~\ref{fig20}, but for the distribution of
    the complete magnitude in SAGES $v_s$ 
    passband, with the median value of $v_s=20.3$ mag.} 
  \label{fig21}
\end{figure}

Figures~\ref{fig20}-\ref{fig21} are the distributions of the complete
magnitude (which defines as the turnover in the magnitude
  histogram) in the SAGES $u_s$ and $v_s$ passband. The median values
are $u_s\sim20.4$ mag and $v_s\sim20.3$ mag.

It is also possible that the photometry table does not reveal
entries for an object in an image, which is clearly visible in the
image in question. Unless SE overlooked the object, given the
parameters we chose, this should only happen when SE extracted two
objects in this image while they are 
considered children of a single parent object for this filter. If all
images within one filter show multiple objects for what is taken to
be a single merged object, all measurements for this object will be
excluded from the photometry table and hence no distilled summary
photometry for this filter shall be included in the master table. 

\subsection{Data Access of SAGES}

Finally we have obtained a master catalog (Please see
Table~\ref{t5.tab}) for the SAGES DR1, which includes 93,293,120
items in $u_s$ band and 91,185,024 items in $v_s$ band. After
combing with PS1, the numbers are reduced to 48,553,987 in our
detection $u_s$ or $v_s$ band.
The SAGES DR1 dataset will be available to the community through the 
China-VO platform and the National Astronomical Data Center (NADC) 
https://nadc.china-vo.org. The release can be browsed at the website, 
either in color or in each SAGES per passband.  

The catalog structure includes the following:

1. The cross-matching to external catalogs now includes SAGES
DR1 and PS1 DR1. Further, in the DR2 version, it will include the 
information of LAMOST and Gaia DR3, as well as 2MASS, AllWISE, GALEX, 
and UCAC4. The LAMOST catalog contains the stellar parameters of gravity, 
metallicity and temperature, while the Gaia DR3 contains the photometry of 
G-band, GB, GR band as well as the parallax and other information.

2. For the SAGES $u_s$ and $v_s$ passband, the photometry table also 
includes a column: if the sources have been measured 1, 2 or 4 times for the
observations, as the overlapped with the adjacent fields. We only
provide the combined magnitude in DR1. Thus for the exposure of 2 or 4
times, the photometry uncertainties are calculated as the variations
of the magnitudes for different times. However, in the next version, 
the individual measurements of photometry may also be released.

\pagestyle{empty}
\begin{deluxetable}{lrr}
  \rotate
  \tablecolumns{3} \tablewidth{0pc} \tablecaption{The item number for
    $u_s$ and $v_s$ passbands for combination themselves and combination with PS1.
    \label{t5.tab}}
  \tablehead{
    \colhead{Parameters} & \colhead{$u_s$-band} & \colhead{$v_s$-band}}
  \startdata
Total detection in $u_s$ or $v_s$ & 98,542,704 & \\
Before combination of $u_s$ or $v_s$  &	  93,293,120 & 91,185,024 \\
	   Combined with PS1 & 48,553,987 &  \\
  \enddata
\end{deluxetable}

\section{Examples of Science Cases}
\label{sci.sec}

The sciences that can be done with a comprehensive survey such as the
SAGES cannot be fully described or predicted. In the following, we outline 
a series of scientific goals that may have a high impact in addressing:

1. Metallicity distribution of the Milky Way halo

The observed metallicity distribution of the Galactic halo can not only help us 
understand its structure, as sub-structures in the halo often stand out in the space 
of metallicity, but also can provide the clue to processes involved in its formation history 
\citep{FN2013book}. Especially, objects in the low-metallicity tail of
the Milky Way metallicity distribution function (MDF) provide a unique
observational window onto the time very shortly after the Big
Bang. They provide key insights into the very beginning of Galactic chemical evolution 
(e.g., see \citealt{Salvadori2010MNRAS} for a typical example). 
Galactic MDFs obtained from the spectroscopic survey data usually suffer complicated 
selection function, whereas photometric survey projects provide a unique chance to 
solve the problem, e.g., recent experiments with the survey data from the SkyMapper 
survey \citep{Youakim2020MNRAS}, the Pristine survey \citep{Chiti2021ApJL}, etc. 
With the large sky coverage and the metallicity sensitive-filter system, the SAGES 
data would be able to provide the best MDF of the Milky Way halo  in the northern 
sky covering the very metal-poor tail, which furthermore results in the most 
complete halo MDF when combining with the southern data from SkyMapper, etc.

2. Searching for stars with the lowest metallicities

Stellar and supernova nucleosynthesis in the first few billion years of the 
cosmic history have set the scene for early structure formation in the Universe, 
while little is known about their nature \citep{BY2011,Karlsson2013RvMP}. 
Since it is not yet possible to directly observe processes of metal enrichment at 
high redshifts, the only observational probe of the metal enrichment sources in the 
early Universe has been chemical signatures retained in the atmosphere of nearby 
long-lived metal-poor stars in the Milky Way and nearby dwarf galaxies. Stars with 
the lowest iron abundances such as extremely metal-poor (EMP) stars with 
$[\rm{Fe/H}]< -3$ are commonly considered to be objects retaining chemical 
signatures of the Pop III nucleosynthesis \citep{BC2005ARAA,FN2015ARAA}.  
With the robust metallicity estimated from the SAGES data, it would enable us to 
carry out the largest scale of searching project for EMP stars and stars with even 
lower metallicities in the northern sky, which would provide a valuable candidate 
database for medium-to-high-resolution follow-up observations. The resulting 
database of Galactic stars with the lowest metallicities can also be used as an 
important probe of early chemical evolution by comparing with chemical evolution 
models \citep[e.g.][]{koba2020}.
 
4. Determine the shape and distribution of dark matter halo of the
Milky Way

The Lambda Cold Dark Matter ($\Lambda$CDM) cosmological model predicts the
existence of dark matter (DM) halos surrounding the Milky Way (MW) and 
external galaxies. Although to first order the spherically
symmetric Navarro–Frenk–White (NFW) profile \citep{nfw97} can
provide a good approximation to the shape of the DM halos, the first
numerical N-body simulations \citep{frenk88,dc91,warr,cola96} found
the shape of the DM halos to be triaxial, and subsequent works
\citep[e.g.,][]{js02,bs05,hay07,vc11}  confirmed these results.  
With SAGES data, candidates blue horizontal branch (BHB) stars can be 
isolated by a machine learning approach through the application of ANN 
combined  with color-color diagram. Since the absolute magnitude of a 
BHB star is relatively stable,  the distance is easier to estimate. 
Then we can examine the number density of the inner halo with our BHB 
stars, which could constrain the shape of the halo.

5. Three-dimensional Distribution of Interstellar Dust in the
Milky Way (north sky)

Extinction and reddening by interstellar dust grains pose a
serious obstacle for the study of the structure and stellar
populations of the Milky Way galaxy. In order to obtain the intrinsic
luminosities or colors of the observed objects, one needs to correct
for the dust extinction and reddening. Extinction maps are useful
tools for this purpose. The traditional two-dimensional (2D)
extinction maps, including those from dust emission in far-infrared
(IR) \citep[hereafter SFD]{sfd98}, far-infrared combined with
microwave \citep{plank14}, and those derived from optical and near-IR
stellar photometry \citep[e.g.,][]{schla10,maje11,nide12,gon12,gon18},
give only the total or an average extinction for a given line 
of sight and therefore do not deliver information on the dust
distribution as a function of distance. This can be particularly
problematic for Galactic objects at a finite distance, especially
those in the disk \citep{guo21}. The Rayleigh-Jeans Color Excess
  (RJCE) method of \citet{maje11} gives the extinction on a
  star-by-star basis, so can be combined with the distance to
  individual stars in a similar fashion. The RJCE method is perhaps
  less accurate because it assumes that all stars have the same color
  in H-4.5 rather than getting the $T_{\rm eff}$ from these passbands,
  but that's a different disadvantage. In \citet{nide12} the RJCE
  results for individual stars are explicitly averaged for a 2-D map.  
With $\alpha_w-\alpha_n$ color, the relation of $T_{\rm eff}$ and 
$\alpha_w-\alpha_n$, the relation of $T_{\rm eff}$ and $g-i$, the 
e(g-i) can be estimated for each star, which is an independent method 
to obtain the extinction of all SAGES stars. Combined with the 
distance of each stars, we can construct the three-dimensional 
distribution of interstellar dust in the northern sky of the Milky Way, 
which will provide another more accurate dust map for the 
astronomical community.

6. Search and identifications for WD candidates 
 
White dwarfs (WDs) are the final stage for the evolution of the majority 
of low- and medium-mass stars with initial masses $< 8M_\odot$.  The 
evolution of  WDs is dominated by a well-understood cooling process 
\citep{fon01,sala00}; because of no fusion reaction. 
WDs are powerful tools with applications in areas of astronomy, such as 
cosmochronology \citep[see,][]{fon01}; constraints on the local star 
formation rate and history of the Galactic disk \citep{krz09}; 
Initial-Final Mass Relation (IFMR) \citep{zhaojk12};
exoplanetary science \citep[see, e.g.,][]{holl18}.  With the 
spectroscopic survey \citep[SDSS, ][]{klei13}; \citep[LAMOST,][]{zhao13}, 
etc. and photometric survey \citep[such as Gaia,][]{gefu21}, 
more and more WDs have been identified.  SAGES is the new data source to 
search WDs candidates. By color-color diagram $g-i$ vs. $u-g$, the  WDs can 
be selected with high confidence in SAGES (Li et al. in preparation). 
Then, the complete WDs sample with 100 pc will be constructed to set up 
the luminosity function, estimate the age of those sample etc. At the 
mean time, new filters (u,v) of SAGES can be used to fit SED, which 
provides more information for WDs population.

7. The substructures of the Milky Way

Large photometric surveys such as 2MASS, SDSS, and Pan-STARRS have revealed 
much tumult at the disc–halo interface through the discovery of
numerous streams and cloud-like structures about the Galactic
mid-plane out to large latitudes ($b<$40 degree) \citep{newb02,sla14}. 
SAGES can provide more accurate stellar parameters. Thus, tracers 
such as RGB stars, BHB stars, etc are easier to separate. With different 
tracer samples, we try to search new streams with multiple methods, such 
as match-filter and integral of motion with the assumption of radial 
velocity distribution. We expect that several new streams in the relatively 
close inner halo can be detected with SAGES data. Also, for known 
substructures, we could identify their member candidates with SAGES data. 
Then their chemical properties will be obtained, which could provide 
constraints for their progenitors.

8. Other science cases, e.g., variable stars, QSOs

In recent years, time domain sciences have become more and more important. 
Although for the SAGES, we only obtain a single exposure for a certain field, we
still can find the sources which have luminosity changes if  
combining with other observations in a similar passband, i.e., SDSS and 
SCUSS. A lot of observations suggest that active galactic nuclei (AGNs) and
Quasi-Stellar Objects (QSOs) exhibit brightness changes in the optical bands. 
\citet{rum2018} presented a sample of $\sim1000$ extreme variability quasars 
(EVQs) with a maximum magnitude change in g-band over one mag with the 
observations of SDSS and Dark Energy Survey (DES) \citep{des05} imaging 
survey. Thus we can compare the $u_s$/$v_s$ band observations of SAGES with 
$u$-band observations of the SDSS and SCUSS to find the variable sources, 
i.e., the QSOs and the variable stars, through the magnitude and color 
transformations.

\section{Future Plans and Data Releases}
\label{fut.sec}

Since March 2015, observations of NOWT of XAO have
focused on the $gri$ passband for the survey, data of which will be
released in the further. Also the observations of the Xuyi 1-m
telescope are still going on and we will release that part when the
observations of DDO51 and H$\alpha$ passband have been finished.

The next data release of SAGES will contain more images and better sky
coverage, as well as co-added sky tiles, where we homogenize the PSFs
of images and then reregister and co-add them within filters. In the future, 
we will also carry out source-finding on co-added frames,
which will provide us deeper detection than now; presently, our
completeness is limited by detection in individual images even though
the distilled photometry has relatively low errors due to the
combination of all good detection into distilled
magnitudes. Forced-position photometry shall become possible as well 
at that time.  

Irrespective of co-added frames, we aim to include PSF magnitudes
that are based on two-dimensional PSF-fitting instead of 1D growth
curves, and are thus more reliable in crowded fields or generally for
objects with close neighbors. 

We plan to process enhancements such as astrometry tied to Gaia DR3 as
a reference frame and better fitting of electronic interference and
CCD bias, especially in areas covered by large galaxies and extended
nebulae, where at present the bias is incorrect, causing excess
noise and over-subtraction of the background. This is relevant for the
creation of high-quality co-added images of galaxies and accurate SEDs
of large galaxies.

Finally, the photometric calibration will also be upgraded in the next
data release.

\acknowledgements 

This study is supported by the National Natural Science Foundation of China
(NSFC) under grant Nos.11988101, 11973048, 11973049, 1222305,
11927804, 11890694 and 11873052, and the National Key R\&D Program of
China, grant no. 2019YFA0405500. This work is also supported by the
GHfund A (202202018107).  
We acknowledge the support from the 2m Chinese Space Station Telescope 
project CMS-CSST-2021-B05. Supported by the Space debris and NEO
research project (Nos.KJSP2020020204 \& KJSP2020020102) and Minor
Planet Foundation.  We thank the staff of University of Arizona and
mountain operation team of Steward Observatory: Bill Wood, Michael
Lesser, Olszewski, Edward W., Joe Hoscheidt, Gary Rosenbaum, Jeff
Rill, Richard Green, etc, for the help of observations. We thank
Prof. Michael S. Bessell for useful discussion and Dr. James E. Wicker
for professional language revision.

\appendix          

\label{lastpage}
\end{document}